\documentclass[12pt]{iopart}
\usepackage{epsfig}  
\begin{document}

\title[High Energy Neutrino Astronomy: Inclined Air Showers]
{Neutrino Detection with Inclined Air Showers}

\author{Enrique Zas  
\footnote[3]{(zas@fpaxp1.usc.es)}}

\address{Departamento de F\'\i sica de Part\'\i culas, Facultad de F\'\i sica, 
Campus Sur, Universidad de Santiago, 15706 Santiago de Compostela, Spain, and}

\address{Kavli Institue for Cosmological Physics, University of Chicago, 
5640 South Ellis Av., Chicago, IL 60637, US}

\begin{abstract}

The possibilities of detecting high energy neutrinos through inclined showers
produced in the atmosphere are addressed with an emphasis on the detection 
of air showers by arrays of particle detectors. 
Rates of inclined showers produced by both down-going neutrino interactions 
and by up-coming $\tau$ decays from earth-skimming neutrinos as a function of 
shower energy are calculated with analytical methods using two sample 
neutrino fluxes with different spectral indices. 
The relative contributions from different flavors and charged, neutral current
and resonant interactions are compared for down-going neutrinos interacting in 
the atmosphere. 
No detailed description of detectors is attempted but rough energy thresholds 
are implemented to establish the ranges of energies which are more suitable 
for neutrino detection through inclined showers. 
Down-going and up-coming rates are compared. 

\end{abstract}

%Uncomment for PACS numbers title message
%\pacs{00.00, 20.00, 42.10}

% Uncomment for Submitted to journal title message

%\submitto{\JPA}

% Comment out if separate title page not required
\maketitle

\section{Introduction}

The observation of Ultra High Energy Cosmic Rays (UHE CR) of energy 
$\sim 10^{20}~$eV\cite{Bird,Hayashida:1994hb,AGASA20,HiRes20} has 
stimulated much theoretical and experimental activity in the field of 
Astroparticle Physics. 
These cosmic rays are are expected to be suppressed by interactions with 
the Cosmic Microwave Background (CMB) radiation, what is known 
as the Greisen Zatsepin Kuzmin (GZK) cutoff\cite{GZK,Zatsepin:1966jv}. 
%and prospects for the construction of a giant air shower array for their 
%study, the Pierre Auger project\cite{Auger}, 
Although many mysteries remain to be solved \cite{naganowatson}, we know it is 
virtually impossible to produce these energetic particles without
associated fluxes of gamma rays and neutrinos from pion decays. 
The gamma rays pair produce in the background photon fields and 
the electrons loose energy through synchrotron emission in a 
``cosmic cascading'' process which degrades their energy down to the 100-MeV 
region. Neutrinos travel undisturbed carrying a ``footprint'' of the 
production model. 
It is remarkable that the diffuse gamma rays in the 100-MeV region, 
UHE neutrinos and UHE CR are deeply related. Their precise measurements, 
crucial for unraveling the mysterious origin of the highest energy 
particles in the Universe, must be considered part of the same priority. 

High energy Cosmic Ray production mechanisms are basically of two 
types, namely acceleration models and ``top-down'' scenarios 
\cite{bottom-up}. Neutrino fluxes in the EeV range are 
difficult to avoid but their fluxes are uncertain \cite{Nureview}. 
In models that accelerate protons and nuclei, 
pions are believed to be produced by cosmic ray interactions with matter or 
radiation at the source. 
In ``top-down'' scenarios protons and neutrons are produced from quark and 
gluon fragmentation, a mechanism which is known to produce about 30 times more 
pions than nucleons. 
Furthermore protons and nuclei also produce pions in their unavoidable 
interactions responsible for the GZK cutoff 
which decay to produce the {\sl cosmological neutrinos} \cite{GZKnu}. 
Cosmological neutrinos could dominate in some acceleration models 
but in top-down scenarios they should be well below the neutrinos from the 
decays of the fragmented pions.  
The UHE proton to neutrino flux ratio is thought to carry 
important information concerning the origin of UHE CR. 

High energy neutrinos produce extensive air showers and can be detected 
by the very same detectors that measure the cosmic ray spectrum. 
The main challenge lies in separating showers initiated by neutrinos by those 
induced by regular cosmic rays. 
It was already suggested in the 1960's that this could be done at high zenith 
angles~\cite{berez} because the atmosphere slant depth is quite large. 
%the electromagnetic component of deeply inclined 
Neutrinos, having very small cross sections, can interact at any point 
along their trajectories while 
most of the air showers induced by protons, nuclei or photons are absorbed 
before reaching ground level. 
The signature for neutrino events is thus inclined showers that interact 
deep in the atmosphere.  
%have a large electron and photon component, 

%{\bf Other Work}
Inclined showers were first observed in the 60's by several groups 
\cite{Tokyo,HP60s}.  
The observed rate for shower sizes between 10$^{3}$-10$^{5}$ particles 
detected in the Tokyo (INS) array \cite{Tokyo} is however consistent with hard 
processes induced by energetic muons that interact deep into the atmosphere 
\cite{Kiraly,tokyobremss,hsvum,ParenteShoup}. For shower 
sizes above 10$^{5}$ particles, AKENO has published an upper bound on muon 
poor air showers at zenith angles greater than 60$^{0}$ \cite{AGASA}. 
These observations have been 
very useful for establishing bounds on models for high energy diffuse neutrino 
fluxes at the Earth \cite{hawaii,blancoPRL}. 
Inclined showers were also detected in Haverah Park (HP), a $12~$km$^2$ air 
shower array in the UK. The largest events detected by Haverah Park were 
actually coming at very large zenith angles \cite{HP60s}. 
The inclined data set was recently studied and shown to be 
consistent with hadronic origin \cite{Ave03}. 
When the Auger project was conceived 
as the largest and most accurate air shower detector \cite{Auger}, it
became clear that it would have a competitive acceptance for inclined showers 
induced by neutrinos compared to devoted neutrino experiments in planning
or construction \cite{ParenteZas,Capelle}. Other experiments have 
been proposed to search for deep inclined showers in the atmosphere 
from satellites at orbital altitudes that allow large volumes of atmosphere
to be monitored to increase the chances of detecting 
neutrinos\cite{euso,owl} or from behind mountains 
~\cite{fargion,Yeh:2004rp,Cao:2004sd}.

Although neutrinos of $\tau$ flavor are heavily suppressed at production, 
neutrino flavor oscillations \cite{oscildiscov}, a maximal $\theta_{23}$
mixing and a 2:1 ratio of $\nu_\mu:\nu_e$ at production, 
lead to approximately equal fluxes of all three flavors 
after propagation over cosmological distances \cite{equalnus}. 
When the $\nu_\tau$ undergoes a charged current interaction it produces a 
$\tau$ which, having a short lifetime, typically decays in flight. 
The process can induce two showers separated by a distance gap that is 
on average proportional to the $\tau$ energy. These events were soon 
discussed in the context of deep underground neutrino telescopes and 
referred to as ``double bang events'' \cite{learned}. 
More recently double bang events have been explored 
for higher energy neutrinos in the atmosphere \cite{AtharParente}. 
As a $\nu_\tau$ propagates through the Earth, charged current interactions 
followed by $\tau$ decays (which always include a lower energy $\nu_\tau$) 
effectively shift the neutrino energy ``regenerating'' the  $\nu_\tau$ flux 
\cite{Saltzberg}. For the higher energy neutrinos these processes can occur 
several times. For neutrinos traversing most of the Earth the emerging flux 
has a characteristic ``pile-up'' at an energy region around the 
PeV scale, except for those entering with high zenith angles. 

It was remarked in 1999~\cite{antoine} that the perspectives of detecting 
$\nu_\tau$ fluxes can increase dramatically for neutrinos that enter the 
Earth surface just below the horizon traversing a relatively small earth
matter depth.  
These neutrinos can undergo a charged current interaction to produce a 
$\tau$ that actually exits the Earth and decays in the atmosphere producing 
an upcoming air shower. 
These ``Earth skimming`` neutrinos can be very effective in producing $\tau$'s 
because of the long $\tau$ range at high energies ($\sim 10$~km) which sets 
the scale of the effective volume. This has turned into a promising field that
has been studied by a number of authors \cite{Fargion:2000iz,bertou,Feng:2001ue,bottai,Gupta:2002ze,Tseng:2003pn,Athar:2003nc,miele,Cao:2004sd}. 
Monte Carlo calculations have been performed for the Surface Array of
the Auger detector \cite{bertou}, for mountain ranges \cite{Cao:2004sd} and 
in general for atmospheric $\tau$ production rates in general \cite{bottai}, 
and there are analytical calculations for the Fluorescence part of the 
Auger experiment\cite{miele} and for HiRes and the Telescope Array 
\cite{Feng:2001ue,Gupta:2002ze}. 
In this article we calculate both down-going and upcoming rates of neutrino 
induced air showers relevant for arrays of particle detectors using 
analytical methods. Comparative studies made by simulation have also been 
discussed by in the context of shower detection from satellites~\cite{bottai}. 
Although the detailed rates in an air shower array are expected to be very 
dependent on detector technicalities we examine the energy spectra of the 
induced showers to compare the down-going neutrino event rate to the 
upcoming rate induced by earth skimming neutrinos.  

\section{EeV Neutrino Detection: Generalities}
%\subsection{Inclined Showers Induced by down-going Neutrinos}

Above 1~PeV the Earth becomes opaque to neutrinos and only down-going or 
earth skimming EeV neutrinos can be detected. The challenge lies in the 
identification of these showers in the background of down-going cosmic 
rays and atmospheric muons. Inclined showers in the atmosphere are expected 
to play a crucial role for the detection of EeV neutrinos. 

The main background for the detection of inclined showers 
produced by neutrinos is due to showers induced by protons and 
nuclei. These are expected to develop high in the atmosphere so that when 
the shower front reaches ground level it has very different properties from 
``ordinary'' vertical showers that are observed in the vertical direction. 
Deep inclined showers induced by neutrinos can develop close to ground 
level so that their shower front resemble that of a typical 
vertical proton cosmic ray shower. 
There is a second type of background to inclined showers induced by neutrinos 
which is due to deep showers induced otherwise, mostly through hard muon 
bremsstrahlung. 
Unlike those produced cosmic rays these cannot be distinguished 
from down-going neutrino induced showers on the basis of the muon interaction 
point because they can also be deep. It is difficult to believe this 
background could be separately identified, alternatively it is thought that 
it sets the detectability limits for neutrino interactions. 

The success in the search for inclined showers induced by neutrinos 
depends crucially on obtaining reliable information on shower 
properties related to the depth of the first interaction. 
Fluorescence detectors actually register the depth development of the 
air shower and the depth of shower maximum, which is closely related to it, 
can be directly obtained. For these detectors upcoming showers, such as 
typical neutrino earth skimming events, can be in principle easily identified. 
In these respects the fluorescence 
technique is most reliable and only good reconstruction accuracy is 
required for neutrino identification. Unfortunately background light
introduces a duty cycle which limits its acceptance both for cosmic ray and 
neutrino
detection. Air shower arrays measure the particle densities as the shower 
front reaches ground level and thus the inference of the depth of the first
interaction is a far more indirect measurement. For close to horizontal events
it is actually not possible to 
distinguish upcoming from down-going events from the arrival times of the 
signals, so that other shower characteristics must be used to identify 
potential neutrino induced events. The study of the 
difference between deep and shallow inclined showers has been the subject 
of much activity in the recent past. It is now believed that there are a 
number of properties of the shower front that can be easily used to
distinguish neutrino induced showers, mostly stemming from the time 
distribution of the shower particles. 
We discuss briefly the basis of these differences. 
However the detailed calculation of the efficiency of a given detector to 
separate deep showers from those induced by protons and nuclei is quite 
a technical issue which lies beyond the scope of this article. 

The neutrino cross sections have a direct impact on the expected shower 
rates. Unfortunately the cross section has to be calculated using quite 
significant extrapolations of data obtained in accelerators and there 
is uncertainty involved. At the end of this 
section we briefly review the basics of the standard model neutrino 
interactions and give the details of the cross sections used for 
completeness. The effects of non-standard model behaviors of the cross 
section can be quite dramatic in the shower rates. These have been discussed 
several times in the literature \cite{Han04} and will not be further 
addressed here. 

%\subsection{Background to Inclined Showers induced by Neutrinos}
\subsection{Inclined Showers from Cosmic Rays}
\label{crbackground}

\begin{figure}[ht]
\centerline{
\mbox{\epsfig{figure=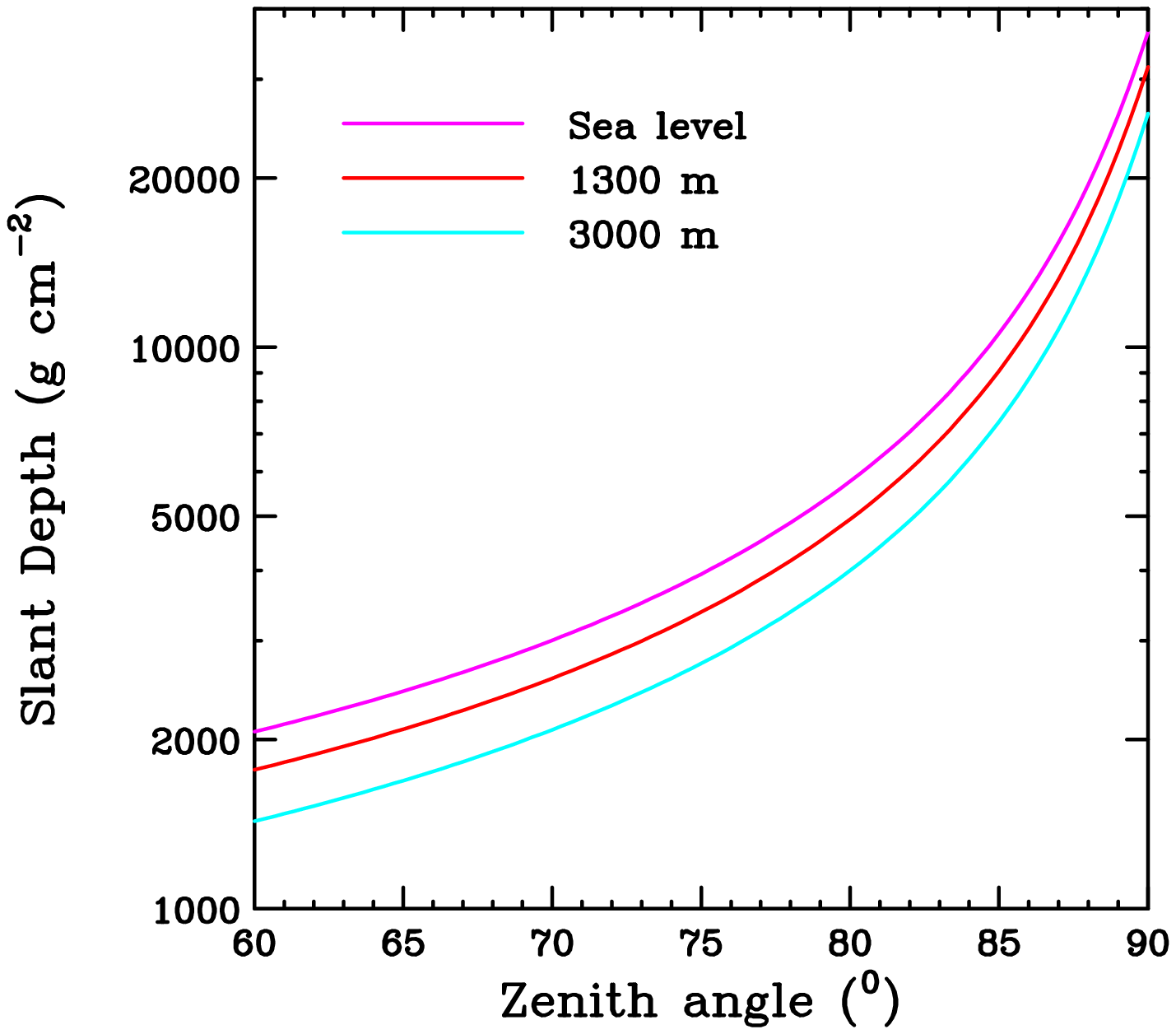,width=8.cm,angle=0},
\epsfig{figure=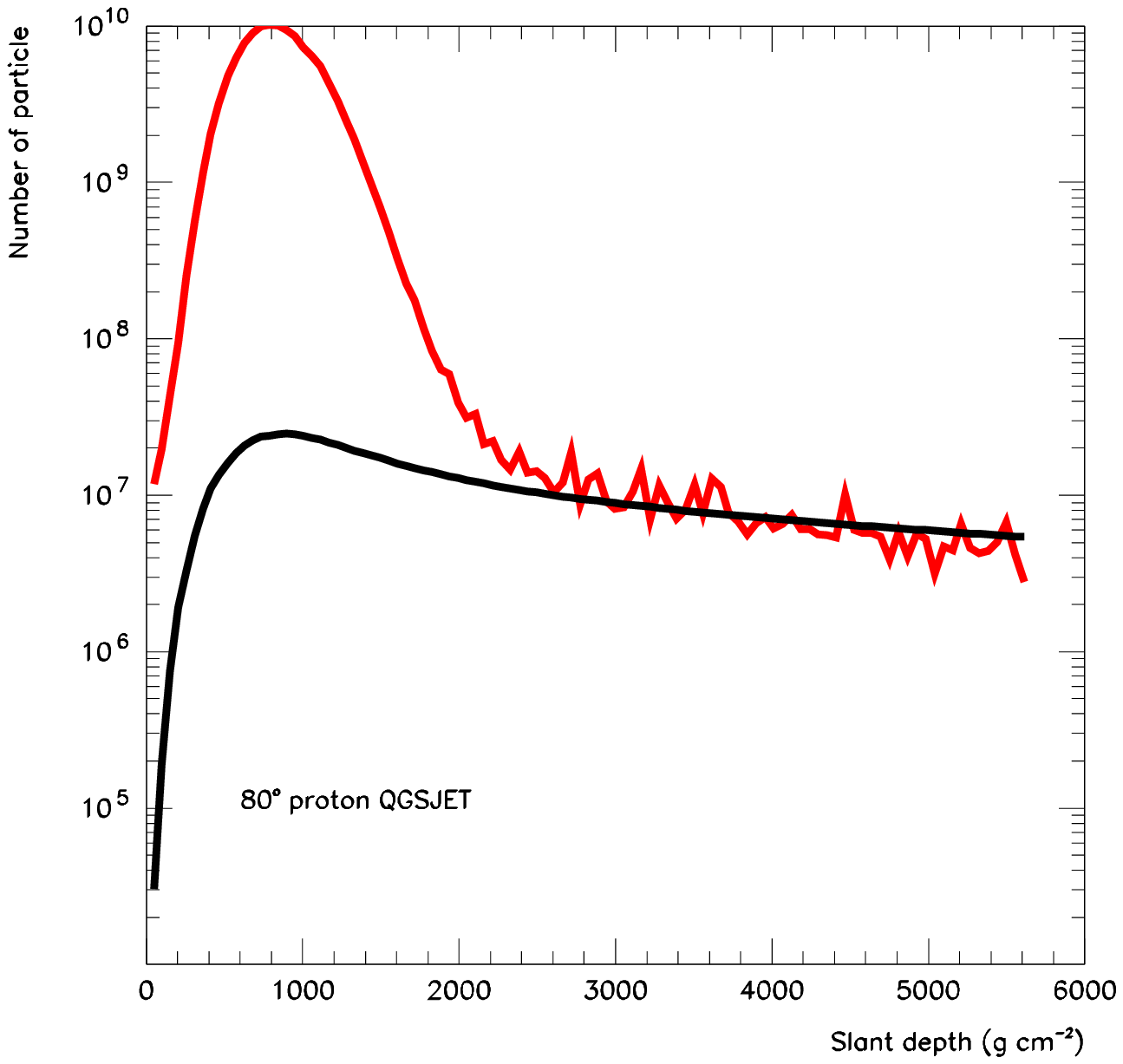,width=8.5cm,angle=0}}
}
\caption{Left: Slant depth as a function of zenith angle for three different 
altitudes: from top to bottom sea level, 1300 and 3000~m. Right: Depth
development of electromagnetic (red) and muonic (black) components of a 
typical shower induced by a proton (qualitatively identical to that induced 
by a nucleus).}
\label{SlantDepth}
\end{figure}

Atmospheric air showers are produced by high energy cosmic rays that interact 
soon after entering the upper part of the atmosphere and they are regularly 
registered by air shower detectors \cite{hires,agasa,EngArray}. 
The atmosphere has just the adequate matter depth ($\sim 1000$~g~cm$^{-2}$) 
so that vertical showers fully develop not too far above ground level, where  
%\footnote{Incidentally this is  why arrays of particle detectors 
%located at the Earth surface are at a near optimal location for maximizing 
%the sensitivity to showers while minimizing the effects of fluctuations.}. 
the shower front %has just gone through shower maximum, and 
contains large numbers of gamma rays, electrons and positrons (the
electromagnetic component) as well as muons in rough proportions 100:6:4:1. 
As the arrival direction of the cosmic ray particles increases in zenith 
angle, $\theta$, the slant depth to ground level rises, 
in proportion to $\sec \theta$ for low zenith angles and with a modified 
behavior for $\theta > 60^\circ$ because of the Earth's curvature. 
The maximum slant depth for a completely horizontal ($60^\circ$) shower at 
sea level is $\sim 36000 ~(2000) $~g~cm$^{-2}$, see Fig.~\ref{SlantDepth}, 
about 36 (2) times larger than for vertical showers. As a result most of the 
electromagnetic component of showers with $\theta > 60^\circ$ is much absorbed 
before reaching ground level. 

The analysis of the Haverah Park inclined data set established the 
cosmic ray EeV background for neutrino detection \cite{rateave98}. 
A fairly general model to describe the inclined showers induced by 
protons, nuclei and photons \cite{HSmodel} was developed which allowed 
the first analysis of EeV data at zenith angles between $60^\circ$ and 
$85^\circ$on on an event by event basis. The observed rate was shown to be 
consistent with a baryonic origin of the cosmic rays and a limit on photon 
abundance above $10^{19}~$eV was set\cite{rateave98,AvePRL,Ave03}. 
The electromagnetic part of cosmic ray showers at high zenith develops and 
gets practically absorbed in the first 2000~g~cm$^{-2}$ and to a very good 
approximation only muons reach the ground. The lower energy muons 
actually decay in flight and can contribute a small electromagnetic 
component that follows closely that of the muons. As a result the average 
energy of the muons reaching ground increases rapidly as the zenith angle 
rises. The magnetic field of the Earth acts as a spectrometer for the
surviving muons separating the negative from the positive in inverse 
proportion to their energy and producing complex density patterns on 
the ground \cite{HSmodel}. The shower front that reaches ground level 
for inclined showers is very different from vertical showers. Inclined 
shower fronts practically only contain energetic muons and their density 
patterns have lost the cylindrical symmetry because of the Earth's 
magnetic field. The degree to which this symmetry is broken depends on 
the relative strength and orientation of the magnetic field with respect to 
shower axis as well as the characteristic distance traveled by the muons
which is governed by the zenith angle \cite{HSmodel}. 

Studies have been made of the time
structure of the shower front induced by inclined cosmic rays and deep 
showers~\cite{billoir}. While the former are very close to a plane front 
deep showers display a front with significant curvature, in similar fashion 
to ordinary vertical showers. This can be characterized by an effective radius 
of curvature that is directly related to the distance at which 
shower maximum is achieved. 
The time structure of the arriving particles has been shown to be very 
different because it is due to muons that are produced very far away. 
The muons do not have much multiple elastic scattering and, as a result, their 
arrival time distribution can be related to a very good approximation only 
to their geometrical path and the kinematic delays associated with the muon 
sub-luminal velocity~\cite{LCB}. At large zenith ordinary 
cosmic rays produce the shower muons very far from ground level and both 
the front curvature and its width are very small. On the contrary 
vertical and deep inclined showers, with large numbers of electrons,
positrons and photons, display both a small radius of curvature because 
shower maximum is relatively close as well as a a large signal dispersion 
in time which increases as the distance to shower axis increases. 
The time structure of the showers detected with the Auger observatory have  
shown clearly these effects \cite{CroninFADC,avetokyo}. 

The inclined shower technique must exploit these differences to search 
for neutrino induced inclined air showers. A simple signature can be used: 
inclined showers that look like ordinary vertical showers with a significant
electromagnetic component. More sophisticated analysis methods can enhance 
the performance of the detectors to search for deep showers for instance
relating the time structure of the muons to the depth development of the
shower \cite{Mccomb:1982nz}. It has been recently shown that it is 
possible to establish the depth of production of the muons in a shower from 
the time distributions of the muons which are far from shower
axis\cite{LCB2}. If this technique can be effectively exploited it could 
allow the selection of deep showers also for showers that have had 
their electromagnetic component absorbed. 

%{\bf MOVE to 2} 
%   *************
%   *************

\subsection{Inclined Showers Produced by Hard Muons}
\label{mubackground}

High energy muons can produce deep air showers by bremsstrahlung,
pair production and nuclear interactions. Although these processes typically
contribute similar amounts to muon energy loss, bremsstrahlung is the hardest 
process and when folded with a steeply falling atmospheric muon spectrum, 
is the most important for producing high energy showers. 
This is fact the dominant mechanism for explaining the observed inclined 
shower rates in the energy range between $100~$TeV and $10~$PeV \cite{hsvum}. 
The high energy muons that are mostly traveling along the shower axis 
must undergo the bremsstrahlung interaction in which a large fraction 
of the muon energy is transferred to the photon. This must be produced 
at about 400~g~cm$^{-2}$ of slant depth before the muon reaches ground 
level, in a relatively narrow range of depth so that the shower reaches its 
maximum close to ground level. Both bremsstrahlung and pair production will 
give rise to pure electromagnetic showers that would hardly contain any muons,
but the hadronic interactions of the muons will produce showers that are of 
hadronic type. 
 
Since atmospheric muons are produced in air showers these energetic 
muons must come from extensive air showers which exceed the muon energy by 
some factor, which is, incidentally, dependent on the nature of the primary 
particle. The hard process itself generates a sub-shower of a fraction of the 
muon energy which develops along shower axis. 
It is even possible that several of these hard muon processes 
combine in a single larger shower. % \cite{EeVMuStanev}. 
The softer muons, which are spread laterally 
in patterns which are governed by the geomagnetic field, could in principle 
also be detected if the primary shower energy is sufficiently high. 

The deep showers induced by these hard processes is thus ``embedded'' in a 
larger shower which may or may not be detectable. If they are detected it 
could in principle be possible to distinguish the deep sub-shower within the 
cosmic ray shower which would clearly signal this kind of processes but this 
seems a rather difficult task. 
If not, it would be impossible to distinguish from a charged current electron 
neutrino shower for instance.  
Fortunately the atmospheric muon flux is very soft and (at
sufficiently high energy) the expected rate of showers from hard muon 
processes is expected to be very small. However the muon bremsstrahlung rate 
is subject to uncertainties, mainly due to the production of prompt muons 
through charmed mesons, a subject that has not been resolved \cite{Gondolo}. 
Inclined showers of shower size between $10^5$ and $10^7$ are expected to 
give valuable information concerning prompt charm production \cite{Gonzalez}. 
Backgrounds of hard muon processes for neutrino detection in the context of
the Auger detector have been discussed in Ref.~\cite{PZaugerMuons}. 

\subsection{The Neutrino Cross Sections}
\label{cross}

The dominant interaction for neutrinos in matter is Deep Inelastic Scattering
(DIS) on nucleons ($\nu + N \rightarrow l + X$, where $N$ stands for a nucleon
and $l$ for a lepton). The cross sections are given in terms of the 
standard structure functions. Within the standard model propagator effects 
associated with the mass of the weak exchange bosons, $M_B$, become 
important at high energies: 
\begin{eqnarray}
%\lefteqn{ 
\label{diffsigma}
\fl {d\sigma \over dy} &=& {G_F^2 m_p E_{\nu} \over \pi} \int_0^1 dx
\left[{M_B^2 \over M_B^2 + 2m_pE_{\nu}xy}\right]^2  \times \\
\nonumber
\fl && \left[\left(1-y-{m_p xy\over 2E_{\nu}}\right)F_2(x,Q^2)+
y^2xF_1(x,Q^2) \pm y\left(1-{y \over 2}\right)xF_3(x,Q^2)\right]
\end{eqnarray}
%
%standard weak interaction cross sections convoluted with 
%
Here $y$ is the fraction of energy transferred to the nucleon in the
laboratory frame, $Q^2$ is minus the square of the 4-momentum transfer, 
$m_p$ is the proton mass and $x$ is defined by the relation $Q^2=2m_pExy$. 
For charged (neutral) current interactions the boson mass is taken to be 
%, $\nu + N \rightarrow l^\pm X$ ($\nu + N \rightarrow \nu X$), 
$M_B=M_W (M_Z)$, the $W^{\pm}(Z^0)$ boson mass. The $\pm$ sign 
in Eq.~\ref{diffsigma} implies 
a $+$ sign for $\nu$ and a $-$ sign for $\bar \nu$.  

$F_{1,2,3}$ are the structure functions which can be expressed in terms of 
universal parton distributions of the nucleons, obtained 
from fits to accelerator data and QCD evolution equations. 
For charged current interactions on isoscalar targets at leading order (LO) 
they are given by: 
\begin{eqnarray} 
F_2^{\nu~cc}(x,Q^2)&=&x(u+d+2s+2b+\overline u+\overline
d+2\overline c+ 2\overline t) \\
xF_3^{\nu~cc}(x,Q^2)&=&x(u+d+2s+2b-\overline u-\overline d-2\overline c-
2\overline t) \\
F_2^{\overline{\nu}~cc}(x,Q^2)&=&x(u+d+2c+2t+\overline u+\overline
d+2\overline s+ 2\overline b) \\
xF_3^{\overline{\nu}~cc}(x,Q^2)&=&x(u+d+2c+2t-\overline u-\overline
d-2\overline s- 2\overline b) \\
F_2(x,Q^2)&=&2xF_1(x,Q^2)
\label{CallanGross}
\end{eqnarray}
where $x$ is to be interpreted as the momentum fraction carried by the
quark distributions ($u,d,s,c,b,t$, their dependence on $x$ and $Q^2$
has been omitted for clarity). For neutral currents the structure functions 
involve different combinations of the parton distributions with neutral 
coupling factors that depend on $z=\sin^2 \theta_W$, where $\theta_W$ is the 
electroweak mixing angle: 
\begin{equation} 
%\nonumber
\fl F_2^{nc}(x,Q^2) =
\left[{8 z^2\over 9}-{2 z\over 3}+{1\over 4}\right] 
x(u+c+t+\overline u+\overline c+ \overline t) +  
 \left[{z^2\over 9}-{z\over 3}+{1\over 4}\right] 
x(d+s+b+\overline d+\overline s+ \overline b) 
\end{equation}
\begin{equation} 
%\nonumber
\fl xF_3^{nc}(x,Q^2) =
    \left[{1\over 4}-{2 z\over 3}\right] 
x(u+c+t-\overline u-\overline c- \overline t) + 
  \left[{1\over 4}-{z\over 3}\right] 
x(d+s+b-\overline d-\overline s- \overline b) 
%F_2^{nc}(x,Q^2) = 2xF_1^{nc}(x,Q^2) && \\ 
\end{equation}
which together with the Callan-Gross relation %~\cite{CallanGross} 
(Eq.\ref{CallanGross}) apply to both $\nu$ and $\overline \nu$. 

Unfortunately the neutrino cross section at the energies of interest 
must be deduced from extrapolations of the parton distributions to 
high $Q^2$ and low $x$ values, where there is no accelerator data. 
As accelerator experiments provide more accurate data on parton 
distributions, new sets of parameterizations are developed which lead 
to somewhat different cross section calculations 
\cite{Reno,Parente,Gandhi:1998ri,Kretzer:2002fr}. 
For energies above $10^6$~GeV charged current interactions are about 2.5 times 
more likely that neutral current interactions. 
Uncertainties can represent up to a factor of two for the total cross 
section at energies around $10^{20}$~eV as estimated in~\cite{Reno:2004cx}. 
Next to leading order \cite{Kretzer:2002fr} and the atomic number of the 
target \cite{castro2} have been shown to imply changes in the cross sections 
which are smaller than the uncertainties due to the extrapolations of the 
structure functions.  
In the work that follows here we will use one of the most recently updated 
sets, CTEQ6 \cite{CTEQ6} for reference. 

This adds an additional interest to neutrino 
detection because neutrino experiments may provide new insights into 
particle physics. 
Besides the obvious role of the neutrino cross section in the actual
interaction that leads to the possible neutrino detection there are 
other more subtle effects. 
The $y$ distribution of the cross section also has an impact on the 
detection rate. The average value obtained with standard parameterizations 
is $<y>\simeq 0.2$ for energies above $10^6~$GeV. 
Since the average energy transfer to the nucleon is very dependent 
on the low $x$ behavior of the parton distributions in a region which 
is not yet accessible to accelerator experiments \cite{castro}, 
it has been pointed out that simultaneous measurements of different neutrino 
channels could provide insight on the low-$x$ behavior of the parton
distributions\cite{alz00}. 
For down-going neutrino induced air showers the interaction 
rate is proportional to the cross section, for Earth skimming upcoming 
showers the cross section acts in the opposite way increasing the neutrino  
absorption in the Earth. A very interesting idea has been discussed that a 
simultaneous measurement of the two rates could lead to 
an indirect measurement of the neutrino cross section at energies well 
beyond accelerator range \cite{WeilerXS}. The relative rates of down-going and
upcoming showers are thus of great relevance from this perspective too. 

Finally we also give an approximate cross section for the Glashow resonance, 
$e~\overline \nu_e \rightarrow W^- \rightarrow \overline l~l'$ 
considering only the first order $s$-channel diagram and neglecting mass 
corrections: 
\begin{equation}
%\lefteqn{ 
{d\sigma \over dy} = {2 G_F^2 m_e E_{\nu} \over \pi} 
\left[{M_W^4 \over (2 m_e E_{\nu} -M_W^2)^2 + M_W^2\Gamma_W^2} \right] 
(1-y)^2 
\end{equation}
where here $y$ is to be taken an the fraction of energy carried by the most 
negative charged lepton, and $\Gamma_W$ is the $W$ boson width. The average
fraction of energy transferred to the most negative lepton is $<y>=0.25$. 
In practice electron cross sections are suppressed by 
the the ratio of the electron to proton mass and this interaction is only 
relevant for $\overline \nu_e$ energy of around $6.4~10^6~$GeV, where it 
dominates over DIS interactions. 
This expression is only adequate for the resonance region. 
The reactions are explicitely shown in Table~\ref{normalizations}. 

\section{Down-going Neutrino Induced Air Showers}

There are multiple channels for producing down-going neutrino induced showers. 
In order of importance first are charged current interactions with 
atmospheric nuclei and then neutral current interactions followed by 
$\overline \nu_e$ resonant interactions with atmospheric electrons. 
Both charged and neutral current neutrino DIS interactions break up  
atmospheric nuclei. The collision debris initiates a hadronic air shower of 
energy $yE_{\nu}$ regardless of the flavor of the original neutrino. 
For charged current $\nu_e$ ($\overline \nu_e$) interactions the emitted 
electron (positron) generates in addition an electromagnetic cascade, 
carrying the remaining energy $(1-y)E_{\nu}$. For charged current $\nu_\tau$ 
interaction the $\tau$ decay can add to the hadronic shower or not depending 
on where it decays. 
Because of the smaller electron mass the interaction of neutrinos with 
atomic electrons is in general suppressed except for the Glashow resonance, 
%\cite{glashow}
$\overline {\nu}_e + e^- \rightarrow W^-$, which dominates
over all other processes at the resonance energy $E_{\bar \nu_e} \sim 6.4 \;
10^6$ GeV\cite{Reno}. The decay of the $W^-$ boson into $q\bar q$ pairs 
dominates because of the six possible final states. If it decays into 
an $e{\bar \nu_e}$ pair, the electron generates a purely electromagnetic 
shower with energy $E \sim 3 \; 10^6$ GeV and if it decays into a 
$\tau{\bar \nu_\tau}$ the shower is produced by $\tau$ decay 
($64\%$ of the times into hadrons and $18\%$ into $e{\bar \nu_e}$). 

\begin{figure}[ht]
\centerline{
\mbox{\epsfig{figure=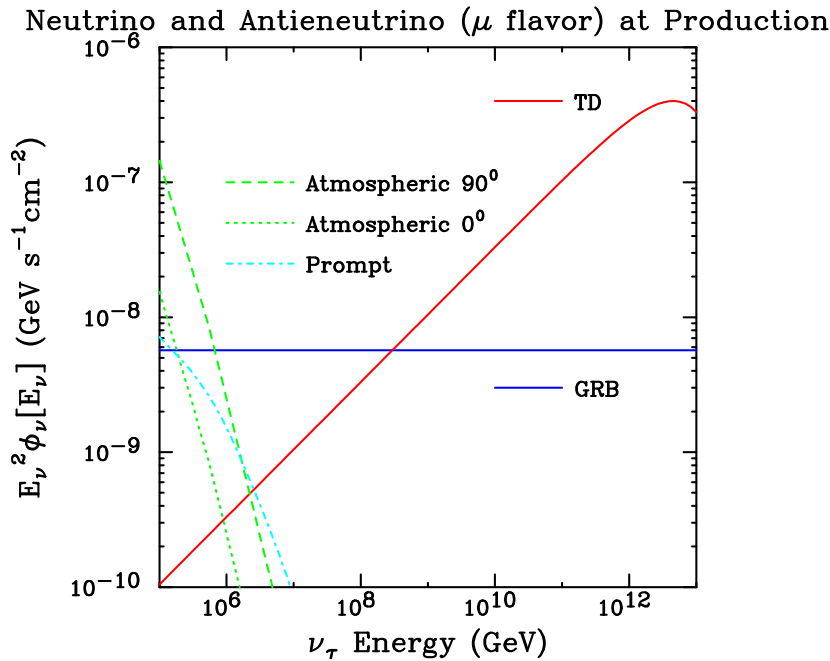,width=11.cm,angle=0}}
}
\caption{Flux of $\nu$ and $\overline \nu$ of each flavor arriving 
to the Earth according to the two parameterizations used for Gamma Ray Burst 
(GRB) and Topological Defect (TD) models, compared to the vertical and 
horizontal atmospheric neutrino fluxes and prompt atmospheric neutrinos 
from charm decays.} 
\label{Fig:fluxes}
\end{figure}

The calculations of the shower rates induced by a differential neutrino flux 
(per unit energy $E_{\nu}$, area, $A$, solid angle, $\Omega$ and time, $t$): 
\begin{equation}
\label{fluxes} 
\phi_{\nu}[E_\nu]= {d^5N_{\nu} \over dE_{\nu} dA d\Omega dt} 
\end{equation}
are based on Ref.\cite{hsvum}. We will used a simple neutrino flux for 
with a constant spectral index of 2 and compatible in the $10^6 -10^8$~GeV 
region with the Gamma Ray Burst 
(GRB) flux given in Ref.~\cite{GRBflux} and one from a characteristic top 
down scenario (TD) 
\footnote{(mass of $X$ particle $M_X=3~10^{14}~$GeV, evolution 
parameter $p=2$ and injection normalization 
$Q_0=10^{-34}~$erg~cm$^{-3}$~s$^{-1}$)}. This last model has been chosen
because it is similar to that used in Ref.~\cite{bertou} and it is consistent 
with experimental data~\cite{TDFlux,Semikoz:2003wv}. They are illustrated in 
Fig.~\ref{Fig:fluxes}. The choices are driven by their different energy 
dependences in the most interesting region between $10^{16}-10^{18}~$eV 
and for comparison with other calculations. 
The total $\nu$ flux is obtained multiplying the $\nu_\mu$ and $\bar \nu_\mu$ 
flux by a factor 1.5 to approximately account for $\nu_e$ production according
to naive counting. 
The basic expression for the calculation 
of the number of showers, $dN_{sh}$, induced by this flux interacting in the 
interval of atmospheric matter depth $dx$, is simply given by: 
\begin{equation}
\label{difShw} 
dN_{sh}=\phi_{\nu}[E_\nu] \left[dA d\Omega dt \right]
dE_\nu ~ N_A dx ~ {d \sigma \over dy} dy 
\end{equation}
where $N_A$ is Avogadro's number. Depending on the interaction the 
differential rates in shower energy have slightly different expressions 
involving the cross sections discussed in Section~\ref{cross}. 
In charged current $\nu_e$ interactions 
all the interaction energy is transferred to the shower: $E_\nu=E_{sh}$ and 
the differential shower rate produced is: 
\begin{equation}
\phi_{sh}[E_{sh}] = {dN_{sh} \over dE_{sh} dA dt d\Omega}= 
\phi_{\nu}[E_{sh}] \sigma^{cc} \int_0^{x_{max}} dx 
\label{nueShRate}
\end{equation}
where $x_{max}(\theta)$ is the slant depth available in the atmosphere 
for the neutrino to interact. 

Unfortunately not all the showers induced by
neutrinos can be detected and distinguished from showers produced by cosmic 
rays. The shower rate that can be identified to be produced by a neutrino 
depends crucially on detector performance and capabilities. We can 
establish a probability for detection ${\cal P}_{det}$, a function that 
necessarily depends on shower energy, arrival direction, 
$d\Omega$, depth at which the shower is produced, $x$, fraction of 
energy transfer, $y$, and impact parameter, ${\vec l}_{\perp}$. 
When this probability is included the shower rate becomes: 
\begin{equation}
\phi_{sh}[E_{sh}] = 
\phi_{\nu}[E_{sh}] \int_0^1 dy {d \sigma^{cc} \over dy}   % \sigma^{cc} 
\int_0^{x_{max}} dx {\cal P}_{det}[y,E_{sh},\Omega,x,{\vec l}_{\perp}]
\label{nueShRate1}
\end{equation}

The above expression can be integrated both in solid angle and in impact
parameter over the surface detector of area $A_{d}$ to give:  
\begin{equation}
\fl \int \int d\Omega dA \phi_{sh}[E_{sh}] = 
\phi_{\nu}[E_{sh}] \int_0^1 dy {d \sigma^{cc} \over dy} 
\left[\int d \Omega \int_{A_d} d^2{\vec l}_{\perp} \int_0^{x_{max}} dx 
{\cal P}^{mix}_{det}[y,E_{sh},\Omega,x,{\vec l}_{\perp}]\right] 
\label{nueShRate2}
\end{equation}
Here ${\cal P}^{mix}_{det}$ refers to the subsequent shower of mixed
electromagnetic and hadronic components respectively carrying fractions 
$(1-y)$ and $y$ of the incident neutrino energy. 
If edge effects are ignored the $d^2 \vec l_\perp$ integral simply factorizes 
as $A_d$ and introduces a ``$\cos \theta$'' factor in the remaining integral. 
The acceptance (as expressed between the large brackets) for the surface 
detector of 
the Pierre Auger Observatory was calculated in Ref.~\cite{Capelle}. 
%${\cal P}_{det}$ was computed 
Expressions for both purely hadronic and purely electromagnetic showers
were obtained integrating in solid angle for $\theta \ge 75^{\circ}$ and 
demanding that the shower intercepts ground level with a significant
electromagnetic component. 

An important contribution to the inclined shower rate induced by neutrinos 
is due to neutral current interactions. 
In these processes the shower energy is simply $E_{sh}=yE_\nu$. 
The result above is modified in the argument 
of the neutrino flux which takes a dependence on $y$, in the $y$ integral 
that must include a $y^{-1}$ factor from the change of variables 
and on ${\cal P}^{had}_{det}$ which now is only for hadronic type 
showers: 
\begin{equation}
\phi_{sh}[E_{sh}] = 
\int_0^1 {dy \over y} \phi_{\nu}[E_{sh}/y] {d \sigma^{nc} \over dy} 
\int_0^{x_{max}} dx {\cal P}^{had}_{det}[E_{sh},\Omega,x,{\vec l}_{\perp}]
\label{nuncShRate2}
\end{equation}
The same expression accounts for charged current $\nu_\mu$ interactions
(replacing $\sigma^{nc}$ by $\sigma^{cc}$)
because the produced $\mu$ is unlikely to decay in the whole atmospheric 
depth unless its energy is below 1~TeV. In the case of $\nu_\tau$ charged 
current interactions, the emerging $\tau$ is likely to decay provided that 
its energy is below $\sim 10^9$~GeV \cite{AtharParente}. 
Since a radiation length in the
atmosphere is of order 300~m, for $\tau$ energies below about
$10^7$~GeV the decay will merge with that from the hadronic debris and 
a result similar to Eq.~\ref{nueShRate2} can be expected. 
For the $\tau$ decay shower 
the fraction of $\tau$ energy that goes into the shower after its decay 
($f$) must also be accounted for. Assuming a constant fraction, 
$E_{sh}=yfE_\nu$, the expression becomes: 
\begin{equation}
\phi_{sh}[E_{sh}] = \int_0^1 {dy \over f~y} {d \sigma^{cc} \over dy} 
\phi_{\nu}\left[{E_{sh} \over f~y} \right] 
\int_0^{x_{max}} dx {\cal P}^{\tau}_{det}[E_{sh},\Omega,x,{\vec l}_{\perp}]
\label{taudecShRate2}
\end{equation}
with ${\cal P}^\tau_{det}$ modified according to the shower details. 
For larger energies the double bang nature of the process will be manifest 
and each of the two showers could be separately detected. 

The calculation proceeds in analogous ways for resonant $W^-$ production with 
$\sigma^{cc}$ replaced by $\sigma^{res}$ in the corresponding expressions. 
When the $W$ boson decays into $q\bar q$ pairs it can be assumed that all the 
energy is channeled after fragmentation into a hadronic shower and 
Eq.~\ref{nueShRate2} applies with a hadronic detection probability function 
${\cal P}_{det}^{had}$ that does nor depend in $y$. As a result the integral 
in $y$ can be trivially performed, what amounts to substituting the integral 
in $y$ for the corresponding total cross section, $\sigma^{res}$. 
When a $e \overline \nu_e$ pair is produced 
Eq.~\ref{nuncShRate2} must be used provided that $y$ (for the resonant 
differential cross section) is taken to be the 
fraction of the neutrino energy carried by the charged lepton in the 
laboratory frame. When the $W$ decays into $\tau \overline \nu_\tau$ pair and 
the shower is produced by $\tau$ decay Eq.~\ref{taudecShRate2} applies. 

%\subsection{Calculation of Shower Rates}
\subsection{Relative Contributions of different Channels}

The precise form of the detection probability ${\cal P}_{det}$ is crucial in 
the calculation of the shower rates. This calculation is 
complex and dependent on technical details of the detector response that must 
include the capabilities to identify showers produced by deep
interactions. To establish some level of comparison with upcoming 
showering events from Earth skimming neutrinos, we make a very simple 
assumption to evaluate it: provided the zenith angle is greater than 
$60^\circ$ all showers that start developing in the second half of the 
slant depth of the atmosphere can in principle be distinguished from 
those produced by cosmic rays. This is not necessarily a conservative
assumption. For air shower arrays the particle detectors must be capable 
of resolving the arrival time and time spread of the signals 
and this is necessarily challenging near 
detector threshold. However by 
demanding that the neutrino travels through at least the first 
$\sim 1000~$g~cm$^{-2}$ without interacting, because of geometrical 
considerations it can be assumed that the subsequent shower must have 
different properties (such as radius of curvature) from a cosmic shower 
of the same zenith. 
\footnote{We remark here that the atmosphere is split in halves in depth. 
This implies that for a detector 1.3 km above sea level and a zenith angle 
of $60^{\circ}$ the midpoint is 10.5~km away while the typical first 
interaction point for a proton or nucleus is $\sim 40$~km away. 
A $\nu$ shower that develops in the second half is closer to the detector by 
more than a factor of two.} 

\begin{table}
\caption{\label{normalizations} Contributing channels to down-going shower 
rates indicating the neutrino reactions, the corresponding neutrino flavors, 
the normalization factors, the shower type, its fraction of the neutrino 
energy and the relevant expression applied.}
\lineup
\begin{tabular}{@{}ccccccc}
\br
Reaction & Type& $\nu$ flavor & Norm Fact & Shower Type & Energy Frac & Eq. \\
\mr
$\nu + N \rightarrow l^\pm X$ & Charged & $\bar \nu_e ~\nu_e$ & 1. & mixed & 1 
& \ref{nueShRate1} \\
$\nu + N \rightarrow l^\pm X$ & Charged & $\bar \nu_\mu \nu_\mu~ \bar \nu_\tau \nu_\tau$ & 1 & hadronic & $\sim 0.2$ & 
\ref{nuncShRate2} \\
$\nu + N \rightarrow l^\pm X$ & Charged & $\bar \nu_\tau \nu_\tau$ & 1. 
& hadronic & $\sim 0.2$ & \ref{taudecShRate2} \\
$\nu + N \rightarrow \nu X$ & Neutral & all~$\nu \overline \nu$'s & 3. & hadronic & $\sim 0.2$ 
& \ref{nuncShRate2} \\
$\nu + e \rightarrow \overline q q$ & Resonant & $\bar \nu_e$ & $0.5 \times 6=3$ 
& hadronic & 1 &  \ref{nueShRate1} \\
$W^- \rightarrow e \bar \nu_e$ & Resonant & $\bar \nu_e$ & $0.5$ & electromg & 0.25 & \ref{nuncShRate2} \\
$W^- \rightarrow \tau \bar \nu_\tau$ & Resonant & $\bar \nu_e$ & $0.5$ & elmg/had &0.25 & \ref{taudecShRate2} \\
\br
\end{tabular}
\end{table}
We then require that showers at ground level have more than a fixed number of 
electrons and positrons ($N_e$) to be detected. 
This requirement will determine the depths at which 
the neutrino can interact to give a significant signal. We make several 
choices of $N_e=10^5,10^6$ and $10^7$ to simulate different detector 
responses naively. These are arbitrary (a shower of $10^{15}$~eV has about 
$10^6$ particles at shower maximum) and set a fairly abrupt low energy cutoff 
for the calculations that follow. 
By choosing deliberately low values the limits of the  
technique are explored by examining its potential low energy behavior. 
In a realistic calculation this is very dependent on the detector, on the 
geometry, on the lateral distribution of the different particles in the
shower front and on many other aspects which have been ignored. 
However as the energy of the shower rises, the range of depths in which the 
neutrino can interact to be detected increases to eventually reach  
half the atmospheric depth. For high energies, when any detector is 
most likely to detect deep showers, we do not expect this simple calculation 
to be very far off. 

The calculations have been made using the above described methods together 
with parameterizations for the depth development of hadronic \cite{Gaisser} 
and of electromagnetic showers \cite{greisen} to calculate $x_{max}$. 
%We choose the $\nu_\mu + \bar \nu_\mu$ flux for reference so that each 
All neutrino flavors are assumed to have the same flux when reaching the 
Earth after mixing. In Table~\ref{normalizations} different channels 
are summarized, together with the resulting normalization factors 
(relative to each neutrino flavor), the type of shower generated, 
the energy fraction transferred to the shower and the appropriate expression
in which $\cal P$ has to be chosen according to the shower type. 
The grouping results from the combinations of the neutrino flavor(s) involved,
the shower type, the fraction of energy transferred to the shower and the 
cross section. 
The normalization factors are associated to the number of neutrino flavors, 
and in the case of $q \overline q$ production a factor of six is 
included to account for the sum over all final state possibilities, 
(two weak doublets and three colors).    

\begin{figure}[ht]
\centerline{
%\mbox{\epsfig{figure=Down.ps,width=10.cm,angle=-90}}
\mbox{\epsfig{figure=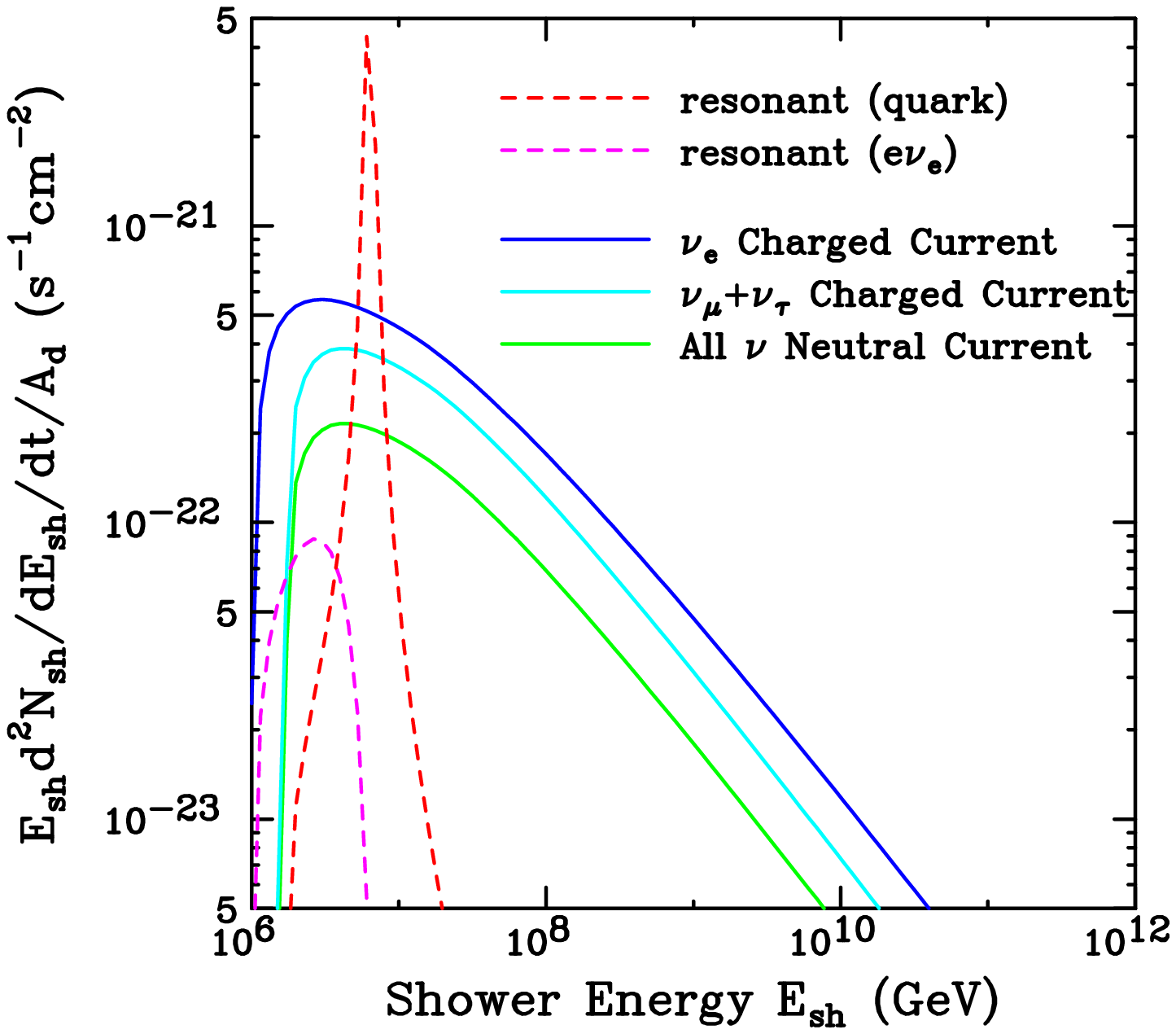,width=9.cm,angle=0},
\epsfig{figure=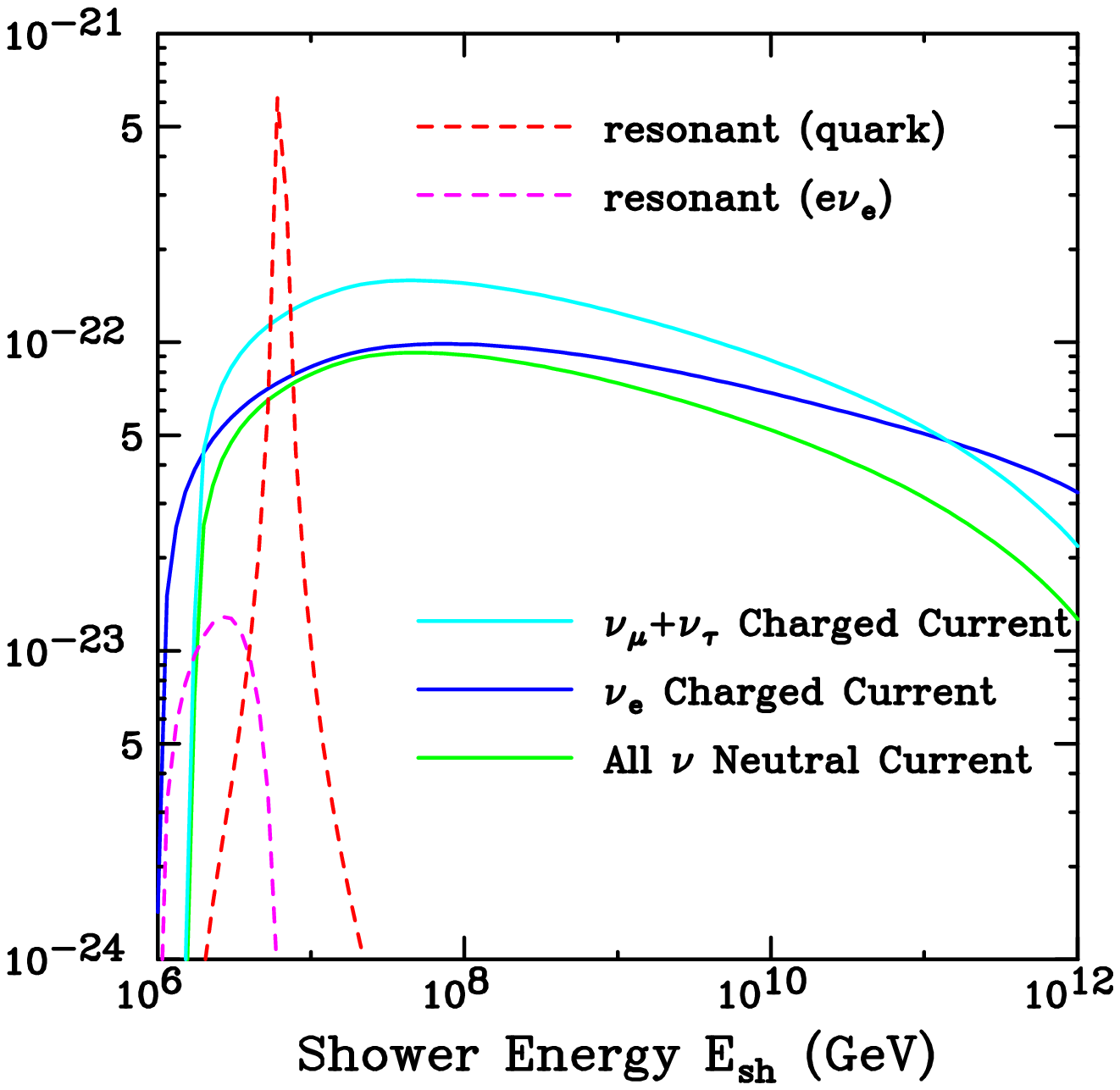,width=8.cm,angle=0}}
}
\caption{Shower rates produced by the main Deep Inelastic Scattering 
processes (full) and resonant $\bar \nu_e e$ interactions (dashed) for the GRB 
model (left) and the TD model (right). DIS interactions are separated in charged 
current $\nu_e + \bar \nu_e$, charged current $\nu_\mu +\bar \nu_\mu +
\nu_\tau +\bar \nu_\tau$ (neglecting $\tau$ decay) and neutral current 
interactions for all neutrino flavors. The $q \overline q$ and 
$e \overline \nu_e$ channels for the Glashow resonance are shown.} 
\label{ShRateGRB}
\end{figure}
The results of the main contributions are compared in Fig.~\ref{ShRateGRB}. 
We have included all the neutral interactions, the $\nu_e,\bar \nu_e$ charged 
current interaction and the $\nu_\mu,\bar \nu_\mu$, the  
$\nu_\tau,\bar \nu_\tau$ charged current interactions neglecting $\tau$ 
decay, (a detailed calculation of this effect will be presented elsewhere 
\cite{PZAuger05}) 
as well as the resonant $q\bar q$ and $e\bar \nu_e$ channels. 
Basically for DIS interactions the calculated shower rate follows closely that
of the incident neutrino flux. The relative contributions of 
each channel depend on its spectral index. This is 
not surprising since the main difference between channels is the fraction 
of energy that goes to the shower. If the $\nu$ spectrum is steep the charged
current electron neutrino interactions dominate because all the neutrino
energy goes to the shower, but for hard fluxes, like the topological defect
models discussed here, the shift in shower energy for neutral current 
channels is compensated by the increase in flavors and hence this channel 
dominates.  

\begin{figure}[ht]
\centerline{
%\mbox{\epsfig{figure=Down.ps,width=10.cm,angle=-90}}
\mbox{\epsfig{figure=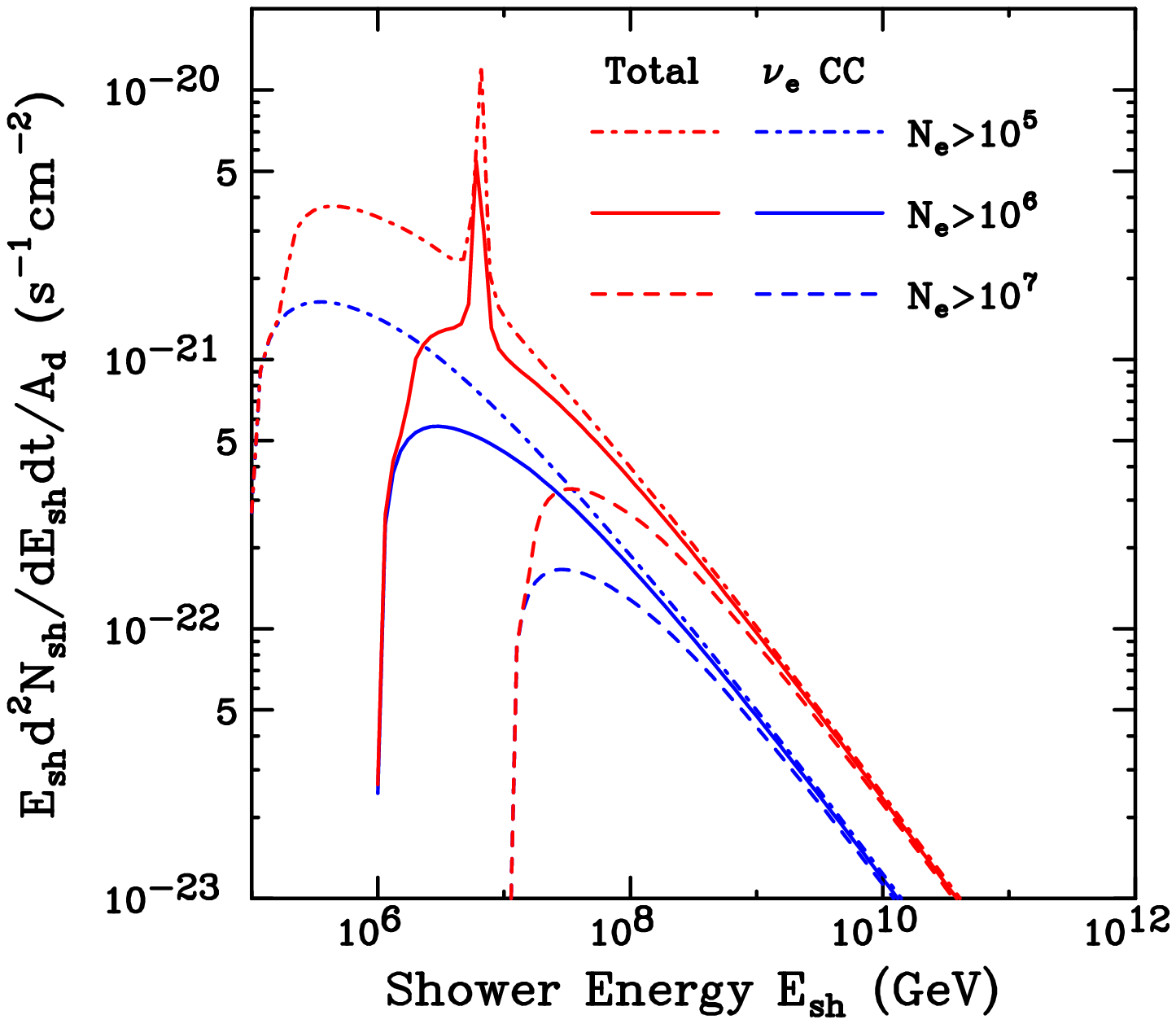,width=9.cm,angle=0},
\epsfig{figure=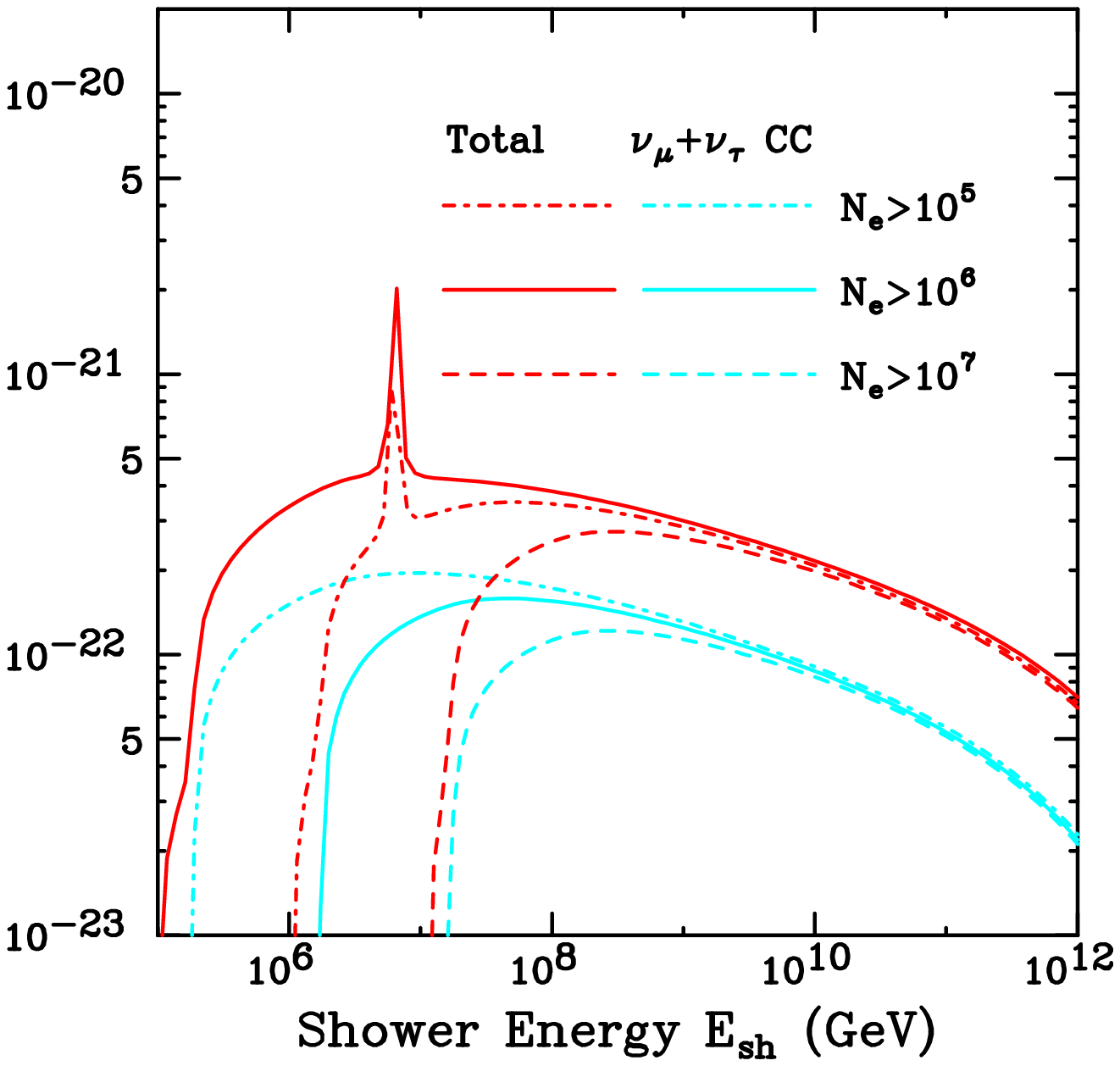,width=8.cm,angle=0}}
}
\caption{Down-going shower rates produced by all channels and flavors added 
up, compared to charged current $\nu_e + \bar \nu_e$ for the GRB model (left) 
and to charged current $\nu_\mu +\bar \nu_\mu + \nu_\tau +\bar \nu_\tau$ for 
the TD model (right). Detector efficiency is roughly estimated requiring 
shower size ($\rho_e$) at ground level to be above nominal threshold values.} 
\label{ShRateTh}
\end{figure}

The total shower rates produced by the discussed DIS and resonant channels 
are plotted for the two fluxes chosen in Fig.~\ref{ShRateTh}, together with 
the dominant DIS channel for each case. In this figure we
also compare the effects of different thresholds as discussed above. The
figure illustrates that the overall shower rate flux is close to a factor of 3
times above that calculated using the dominant neutrino channel only (the
precise factor depends on the spectral shape of the flux in question). The
figure illustrates the dramatic effect that the low energy behavior of an air
shower array can have in the total event rate calculation. Naturally this is 
particularly significant for the GRB model which is much steeper.  
\section{Earth Skimming Neutrinos}

As neutrinos go through the Earth they can undergo charged current 
interactions close to their exit point below the surface of the Earth. 
Under certain circumstances these interactions can produce extensive 
air showers that develop upwards in the atmosphere. 
Only those produced by $\nu_\tau$'s are of significance from the quantitative 
point of view, because they lead to large detectable shower rates in the 
high energy range of interest. This is due to the combined effect of 
the lepton lifetime, its energy loss and the neutrino cross section. 
For $\nu_\tau$ the mechanism can be described in three stages, 
namely the $\nu_\tau$ enters the Earth and propagates through it, 
it then has a charged current interaction just below the surface producing 
a $\tau$ that continues the propagation in matter with considerable 
energy loss and finally the $\tau$ exits the surface and decays in the 
atmosphere inducing an air shower. 
%In the case of $\nu_e$ the produced electron induces a shower as soon 
%as it interacts in the Earth. If this happens sufficiently close to the 
%surface it will partially develop into the atmosphere in a shower that 
%is not too different from that induced by a $\tau$ decay. 

\subsection{Propagation in the Earth}

Neutrinos have a cross section which increases as the neutrino
energy rises. As a flux of neutrinos goes through a given amount of matter 
a characteristic energy scale is selected, the energy at which 
the neutrinos are likely to interact. Above this energy the flux 
is attenuated while below it is practically undisturbed. Both charged 
and neutral current interactions contribute to neutrino attenuation. 
For a neutrino crossing the Earth the amount of matter depth is a rapidly 
varying function depending on impact parameter. If they travel through the 
center, that is they enter with a relatively low zenith angle, 
the matter depth is not much below $6.6~10^{9}$g~cm$^{-2}$ (the depth of 
a diameter). % using the density profile in Ref.~\cite{density}). 
Using the cross section discussed in 
subsection~\ref{cross} this corresponds to the attenuation length of a 
100~TeV neutrino. As the impact point gets further away from the
Earth's center, and both the incidence (zenith) angle and exit (nadir) angle 
(which are the same) approach $90^\circ$, the matter depth 
decreases to zero. At an incidence
zenith of $89^\circ$ the depth has reduced by over a factor 100 to a value 
(which depends on local density) typically between 
$\sim 2-5~10^7$~g~cm$^{-2}$. 
Since the cross section rises with energy approximately as 
$E^{1 \over 3}$ in this range, the characteristic energy scale, at which the 
neutrinos become attenuated changes from 1 PeV to about $10^{19}-10^{20}~$eV 
at $89^\circ$. 
The Earth thus filters the neutrino spectrum in a complex way suppressing 
the high energy part in a rapidly changing way as the zenith angle approaches
the horizontal. 

Besides attenuation there is a regeneration effect. 
Neutral current interactions shift the neutrino energy, transferring part of
the absorbed flux into lower energies. In addition, the $\tau$ ($\mu$) lepton 
produced in charged current interactions can in turn decay producing 
$\nu_\tau$ ($\nu_\mu$). This double sequence also regenerates the neutrino 
flux \cite{Saltzberg}. 
Typically the absorbed high energy $\nu_\tau$ flux gives a $\tau$ which only 
decays when it is below a characteristic energy scale $\sim 10^8$~GeV 
(this is described in subsection~\ref{sec:eloss}). Regeneration is important 
below the smaller of this scale and the energy scale fixed by the absorption. 
For low zenith the regeneration is important in the 1~PeV region 
~\cite{Saltzberg} but for Earth skimming neutrinos regeneration is most
important at about $2~ 10^7$~GeV as shown by simulations \cite{RegDutta}. 
These neutrinos have to interact produce a $\tau$, exit the Earth 
and decay in the atmosphere so that the shower energy is further reduced. 
Since the electron is stable there is no such effect for $\nu_e$ while for 
$\nu_\mu$ the long $\mu$ lifetime results in regeneration at a low energy 
which is not relevant for this work. We will here consider neutral current 
regeneration but will not further discuss the $\nu_\tau$ regeneration. 
Earth skimming rates obtained particularly for the topological defect flux 
will be somewhat higher at around $10^7$~GeV.  

\subsection{Lepton Propagation: Energy loss and Decay}
\label{sec:eloss}

Once a $\nu_\mu$ or a $\nu_\tau$ undergoes a charged current interaction, 
the produced lepton has a probability to decay which increases as it 
loses energy while propagating in matter. The propagation of the 
lepton in the Earth determines the effective volume that is available 
for detection in this channel. 
Following procedures similar to those described in \cite{miele,LCB} we 
calculate the survival probabilities for leptons with continuous energy
losses.  
The survival probability has the following differential equation:  
\begin{equation}
\label{SurvEquat}
 d P_{\rm Surv} = P_{\rm Surv} {dl \over c \beta}{ m c^2 \over E(l)} 
\end{equation}
where $l$ is the distance traveled, $\beta c$ the lepton velocity, $m$ its 
mass, $\tau$ its decay lifetime and $E(l)$ its energy, which decreases as 
$l$ increases. 
 
The energy loss can be expressed by an approximate expression: 
\begin{equation}
\label{Eloss}
 {dE \over d x}= a + b E = {\epsilon + E \over \xi}  
\end{equation}
in which the energy loss per unit depth ($x$) has a constant energy loss 
term associated with ionization ($a$) and a linear term in energy which is due 
to hard processes, bremsstrahlung, pair production and hadronic interactions. 
This is usually expressed in terms of a characteristic grammage, $\xi=1/b$, 
and a critical energy, $\epsilon=a \xi$ above which radiative processes 
dominate. These values are slightly varying with energy and somewhat 
uncertain but they have implications for the $\tau$ production 
rate~\cite{bertou}. We here use the values $\xi=1.25~10^6$~g~cm$^{-2}$ 
and $\epsilon=3000$~GeV consistent with results for the $10^8$~GeV 
region in Ref.~\cite{Dutta:2000hh}. 
This equation can be integrated to give \cite{Lipari}: 
\begin{equation}
\label{E-Eprime}
{E+\epsilon \over E'+\epsilon}= e^{- x / \xi}  % e^{-x \xi^{-1}} 
\end{equation}
which relates $E'$, the initial lepton of energy, to $E$, the lepton energy 
after propagation over depth $x$. 
This expression can be used to express $E$ in Eq.~\ref{SurvEquat} 
in terms of $l$, assuming the lepton propagates at the speed of light in a 
constant density medium. With this substitution Eq.\ref{SurvEquat} can be 
integrated in $x$ to obtain:  
\begin{equation}
\label{SurvProb}
P_{\rm Surv} = \left[ {E \over E'}{E'+\epsilon \over E+ \epsilon}\right]^\eta 
\simeq \left[1-\epsilon \left({1 \over E} - {1 \over E'} \right) \right]^\eta 
\end{equation}
where the second approximate expression holds for $E>>\epsilon$ and the 
exponent, $\eta$, is a constant that depends on the medium density 
$\rho$, on the lepton mass, $m_l$, its decay constant, $\tau_l$, and the 
loss parameters $\xi$ and $\epsilon$: 
\begin{equation}
\label{eta}
\eta = {m_l c^2 ~ \xi \over \epsilon ~ \rho c \tau_l}  
\end{equation}
The value of $\eta$ for muons in 
rock of density $2.65~$g~cm$^{-3}$ is $\sim 3.5~10^{-4}$ which implies that  
decay can be ignored for all cases of interest here (the muon lifetime is 
very long compared to energy loss time). In the atmosphere $\eta \sim 0.66$ 
and the muon can decay in flight. However for the 
energies of interest here the muon exits the atmosphere well before decaying. 
%$E' << \epsilon$ a $50\%$ decay probability 
%implies that the muon must loose about $65 \%$ of its energy. For the 
%$E' >> \epsilon$, $50\%$ decay probability implies 
%that the muon must loose practically all its energy to reach 
%$\sim 1.5 \e. 

For the $\tau$ lepton propagating in standard rock 
$\eta \simeq 3.2~10^4$ is very large. This implies that unless 
the ratio of energies in Eq.~\ref{SurvProb} is very close to one 
the $\tau$ does not survive. We can then approximate Eq.~\ref{SurvProb} as: 
\begin{equation}
\label{SurvProb2}
P_{\rm Surv} \simeq e^{- \eta {1- E/E' \over 1+ E/\epsilon}} 
= e^{- \eta {\epsilon \over E+ \epsilon} {E'- E \over E'}} 
\end{equation}
For the $\tau$ not to decay the absolute value of the exponent must be small 
and the fraction of energy lost by the $\tau$ is limited: 
\begin{equation}
\label{FracEloss}
{E'-E \over E'} < {E+\epsilon \over \eta \epsilon} 
\simeq 3.1~10^{-5} \left[ 1+{E \over \epsilon}\right] 
\end{equation}
For $E$ is below $\epsilon$ ionization losses dominate and the fraction of 
energy loss must be below a very small value. 
When $E >> \epsilon$ this bound increases linearly with $E$, the energy of the 
$\tau$ after propagation, until it becomes 1 when $E \simeq \epsilon \eta$. 
This introduces a new energy scale for the emerging $\tau$: 
$\epsilon \eta \simeq 9~10^{7}~$GeV. 

\begin{figure}[ht]
\centerline{
\mbox{\epsfig{figure=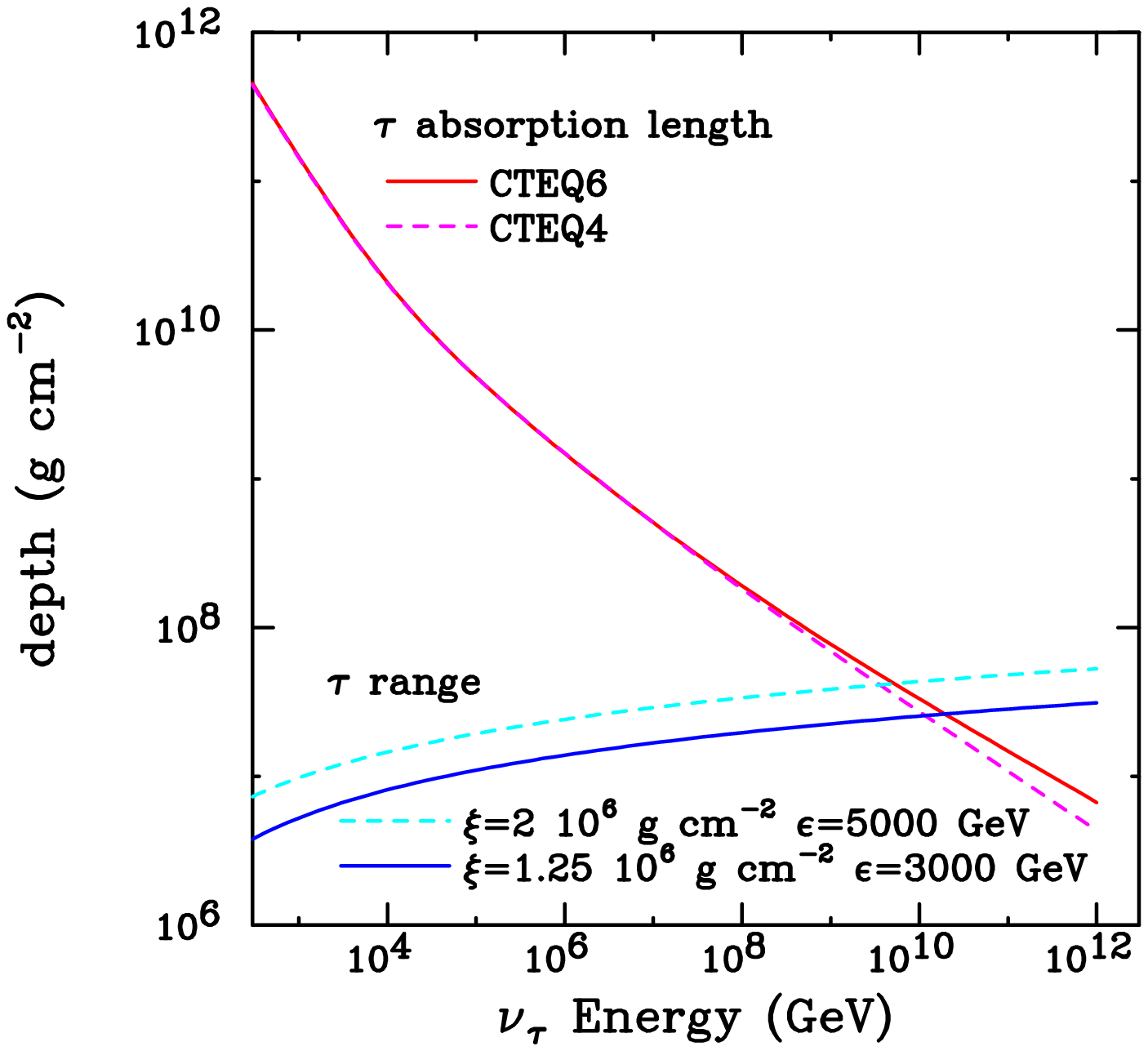,width=9.cm,angle=0},
\epsfig{figure=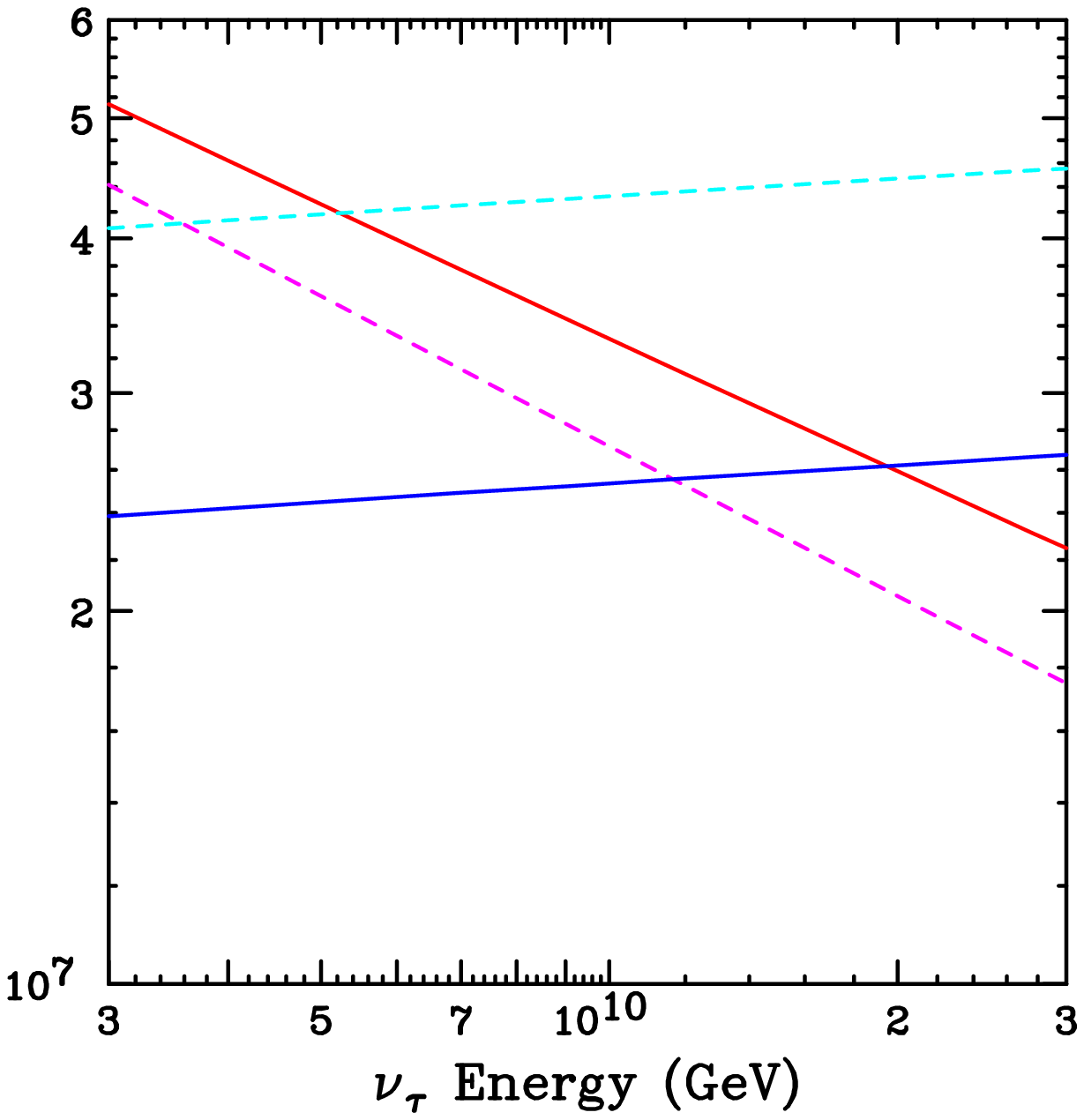,width=8.cm,angle=0}}
}
\caption{Comparison of the $\tau$ range and the $\nu_\tau$ absorption length, 
using two different neutrino cross sections (CTEQ4 and CTEQ6) and and two 
different sets of parameters for $\tau$ energy loss. The right hand side panel
is a blow up of the intersection region.}  
\label{absorb}
\end{figure}
The fraction of energy loss can be related to the $\tau$ range in rock 
through Eq.~\ref{E-Eprime}. Expanding the range for $x << \xi$ we obtain: 
\begin{equation}
\label{E-Eprime:expand}
E'-E \simeq {x \over \xi}[E' + \epsilon] 
\end{equation}
For $E < \epsilon \eta \simeq 9~10^{7}~$GeV the fraction of energy loss 
is below 1 and the above approximation holds. The range increases 
approximately linearly with $E$ and so does the effective volume for 
neutrino detection through Earth skimming neutrinos. 
This is an important energy scale because above it, the effective volume 
for neutrinos ceases to grow linearly with $E$. A second energy scale 
arises because of absorption. 
In Fig.~\ref{absorb} the range at which the survival probability is $0.5$ is 
shown, plotted as a function of $E'$. It is compared to the matter depth at 
which the flux has been reduced by a factor of 2 because of absorption. 
Two different cross sections and two different values of the energy loss
parameters have been used to illustrate the uncertainties involved. 
The crossover energy sets the limiting scale to a value between 
$10^9$-$10^{10}$~GeV above which neutrino absorption suppresses the 
$\tau$ rate. 

%\subsection{Earth Skimming Rate Calculation}
%\subsection{Charged Current Interactions Near the Surface}

\subsection{The Emerging $\tau$ flux}

The calculation of the emerging $\tau$ flux proceeds in a relatively simple 
manner. Neutrinos arriving at a given zenith angle propagate through the 
Earth to a point along the corresponding Earth chord at which they interact. 
The total matter depth along the chord is $x_T(\theta)$ and the depth of the 
interaction point, $x$, is measured along the chord from the exit point. 
We assume that the neutrino flux at $x$ can be expressed as 
$\phi_\nu[E_\nu] e^{-g(x_T-x,E_\nu)}$. The $\tau$ rate produced in an interval
of distance $dx$ is simply obtained convoluting the differential neutrino 
flux at the interaction point with the differential cross section 
$d \sigma^{cc} / dy$ multiplied by the $\tau$ survival probability:
\begin{equation}
\label{difTau} 
{dN_\tau \over dA d\Omega dt}=
\phi[E_\nu] e^{-g(x_T-x,E_\nu)} dE_\nu N_A dx {d \sigma^{cc} \over dy} dy
\left[ {E_\tau \over E'_\tau} 
{E'_\tau + \epsilon \over E_\tau + \epsilon} \right]^\eta 
\end{equation}
where $E'_\tau$ and $E_\tau$ are respectively the $\tau$ energies at the
interaction and exit points and $\rho$ is the matter density at the
interaction point. 

Using Eq.~\ref{Eloss} we can express $dx$ in terms of $dE_\tau$ to get the 
differential $\tau$ rate as the following integral:   
\begin{equation}
\label{Eq:TauFlux}
\fl \phi_\tau[E_\tau,\theta] = {\xi \over E_\tau + \epsilon} N_A 
\int^\infty_{E_\tau} dE_\nu \phi[E_\nu] e^{-g(x_T-x,E_\nu)} 
\int^{y_{max}}_{y_{min}} dy \; \frac{d\sigma^{cc}}{dy} 
\left[ {E_\tau \over E'_\tau}
{E'_\tau + \epsilon \over E_\tau + \epsilon} \right]^\eta 
\end{equation}
The interaction point enters the expression through the function 
$g(x_T-x,E_\nu)$. 
Once $E_\nu$ and $y$ are established, the $\tau$ lepton is produced with 
fixed energy $E'_\tau=E_\nu(1-y)$. If $E_\tau$ is also fixed the 
interaction point $x$ can be expressed in terms of the three variables 
$x(E_\nu,E_\tau,y)$ as: 
\begin{equation}
\label{xinteraction}
x = \xi \ln \left[ {E_\nu(1-y)+\epsilon \over E_\tau+\epsilon} \right] 
\end{equation}
The limits of the $y$ integral in \ref{Eq:TauFlux} are: 
\numparts
\begin{eqnarray}
\label{ylimits}
y_{min}&={1 \over E_\nu} 
\max[E_\nu + \epsilon - (E_\tau + \epsilon) e^{x_T/\xi},0] \\
y_{max}&=1-{E_\tau \over E_\nu} 
%e^{x_T \xi^{-1}},0]
\end{eqnarray}
\endnumparts

\begin{figure}[ht]
\centerline{
%\mbox{\epsfig{figure=Down.ps,width=10.cm,angle=-90}}
\mbox{\epsfig{figure=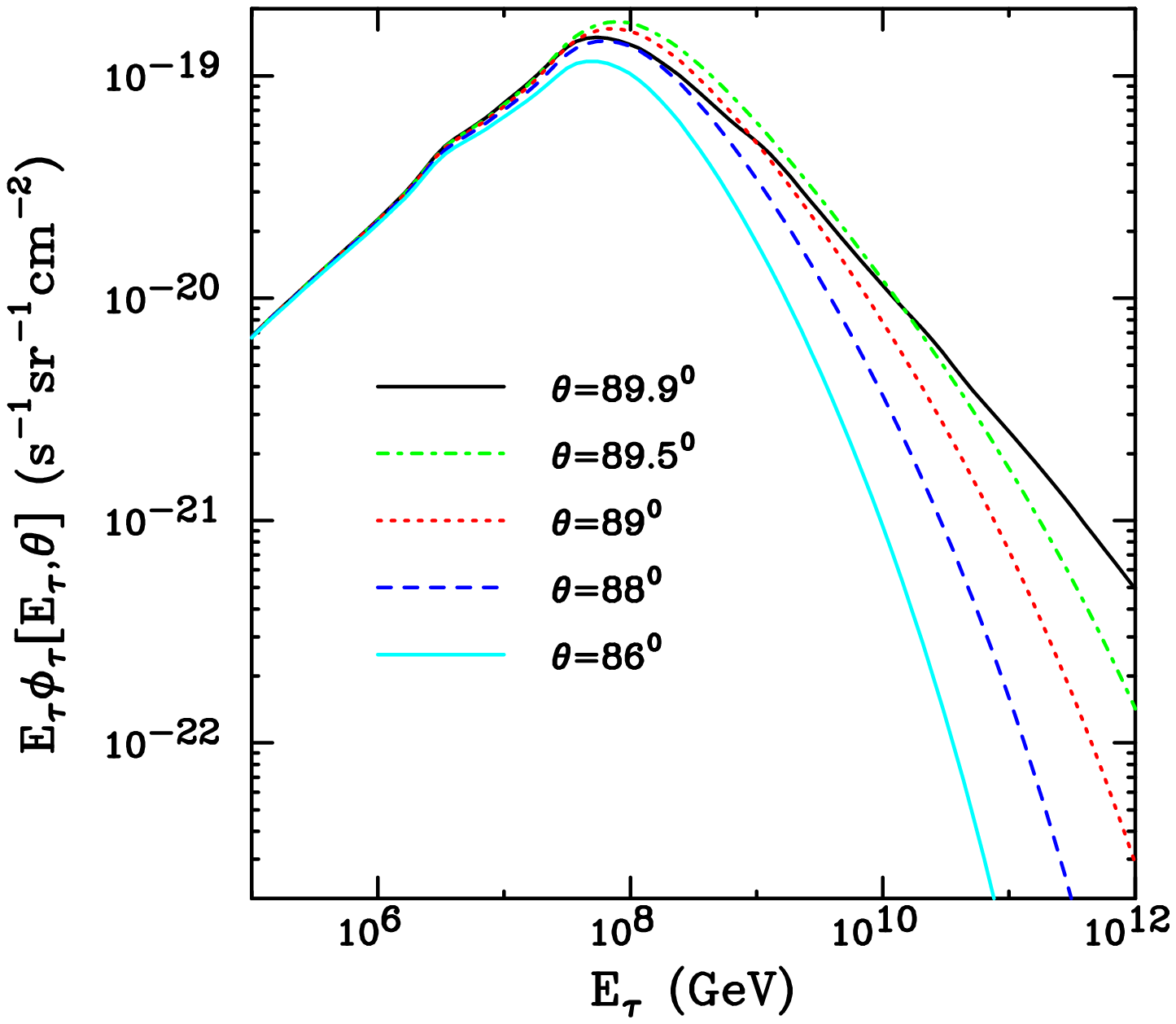,width=9.cm,angle=0}
\epsfig{figure=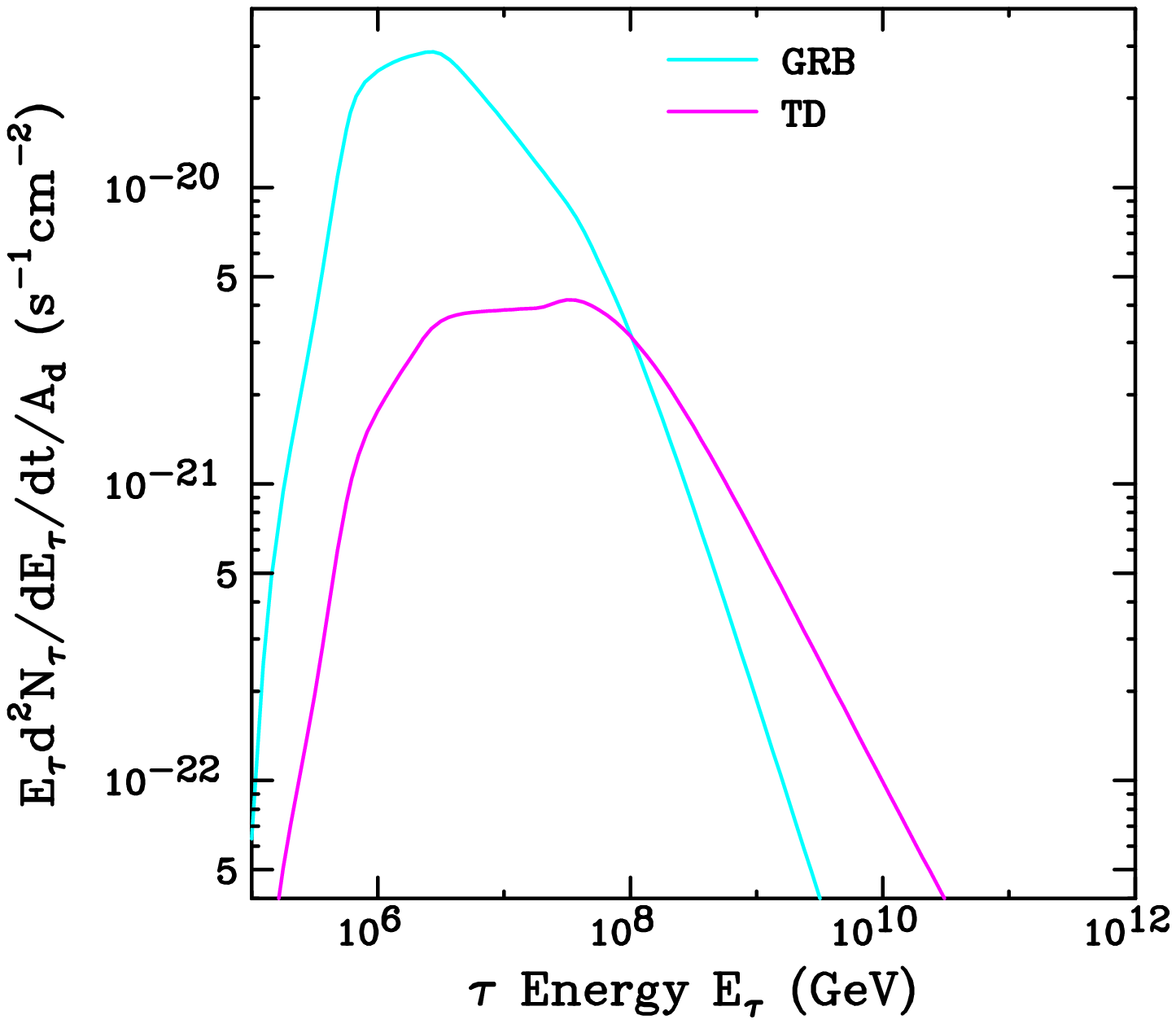,width=8.cm,angle=0}}
}
\caption{Differential $\tau$ spectra produced by earth-skimming neutrinos. 
All $\tau$'s are included and the spectra are differential in $\tau$ 
energy, $E_\tau$. 
Left panel: $\tau$ spectra for the GRB flux at different zenith angles. 
Right panel: $\tau$ spectra after solid angle integration 
for both the TD and GRB fluxes.} 
\label{TauFlux}
\end{figure}
The corresponding fluxes at different zenith close to the horizontal are 
illustrated in Fig~\ref{TauFlux} for the GRB flux used. A strong 
zenith angle dependence is observed which 
%The $\theta$ dependence of 
arises in Eq.~\ref{Eq:TauFlux} from $x_T$, both in the
function $g(\overline x, E_\nu)$ (with $\overline x=x_T-x$) and in the limit 
$y_{max}$. 
To a good approximation the function $g(\bar x, E_\nu)$ can be thought of as 
an attenuation due to the total cross section $\sigma_{tot}$ which is 
effectively reduced by regeneration. If we ignore
regeneration due to the $\tau$ decays in matter we can simply express it as 
$g(\bar x, E_\nu)= (x_T-x) \sigma_{eff}$ where the effective absorption cross
section is given by: 
\begin{equation}
\fl \sigma_{eff}=\sigma_{tot}-\sigma_{reg}=\sigma_{tot}-
 \frac {1} {\phi_{\nu}(E_{\nu})} \; \int_0^1 dy \; 
\phi_{\nu} \left( \frac {E_{\nu}} {1-y} \right) \; 
\frac {d \sigma^{nc}} {dy} \left( \frac {E_{\nu}} {1-y} , y \right)
\end{equation}
and $d \sigma^{nc}/dy$ is the differential neutral current interaction cross
section. It thus depends strongly on the zenith angle. 
We can finally integrate Eq.~\ref{Eq:TauFlux} over solid angle to obtain the 
energy spectrum of the total emerging $\tau$ rate. 
The results for both fluxes are also shown in Fig.~\ref{TauFlux}. 
We note that the bulk of the $\tau$ lepton spectrum is contained within about 
$3~10^5-10^7~$GeV for the prediction from GRB's, while for topological 
defects, which is a harder flux, the range is $\sim 10^6- 3~10^8~$GeV. 

The earth skimming $\tau$ flux as a function of $E_\tau$ 
is a complex result from the competition of an increasing effective 
volume and cross section for the neutrinos and a decreasing solid angle 
because of neutrino absorption. 
This is convoluted with a neutrino flux spectrum which often falls with 
energy like $\sim E_\nu^{ -\gamma}$.
The effective volume increases linearly with $E_\tau$ until the scale of
$9~10^7$~GeV is reached above which the rise is slow. 
The cross section increases with energy as $E^{\sim 1/3}$ for 
$E_\tau > 100~TeV$ while the solid angle decreases with approximately 
the inverse energy dependence. 
If $\gamma > 2$ the dominant part of the 
$\tau$ flux will be for energies at the 100~TeV scale which include all 
directions. But if $\gamma < 2$ the rate will be dominated 
by the turnover scale of $9~10^7$~GeV. For hard spectral indices of order one, 
the dominant part of the flux shifts to the absorption scale $\sim 10^{10}$~eV. 

\subsection{The $\tau$ Decay in the Atmosphere}

Finally the emerging $\tau$ must decay to produce a detectable shower. 
The decay process is mediated by a $W$ boson and always involves a 
$\nu_\tau$ which is irrelevant from the point of view of the developed 
shower. The charge boson couples to both electron and muon lepton pairs  
with probabilities close to $18 \%$. When a $\mu \nu_\mu$ pair is
produced there is no shower, while if a $e \nu_e$ pair is produced about one 
third of the $\tau$ energy is expected to go into an electromagnetic shower. 
The remaining $64 \%$ of the times it couples to a quark antiquark pair that 
fragments 
into hadrons, mostly between one and three hadrons. These showers can be
considered hadronic. Assuming that 
the shower carries all the quark energy we can expect 
about two thirds of the total $\tau$ energy to go into the air shower. 
When all the probabilities are weighted together the average fraction of 
energy carried by the shower is just below $50\%$. 

The $\tau$ travels a distance $l$ in the atmosphere to decay and produce a 
secondary shower that develops to reach shower maximum at a further distance 
$l_{sh}$, which is a slowly varying function of shower energy (of order 10 
radiation lengths or $\sim 3$~km in air for both electromagnetic and hadronic 
showers). The altitude at which the shower reaches shower maximum is crucial 
for its detection and affects different types of detectors in different ways. 
This altitude is a combination of both the decay length and 
the distance to shower maximum ($l+l_{sh}$), 
the nadir angle with which the $\tau$ exits and the curvature of the Earth. 
The differential decay probability is again given by Eq.~\ref{SurvProb}, 
where we can now neglect $\tau$ energy loss which is very small. The
survival probability after traveling distance $l$ is simply: 
\begin{equation}
\label{DiffSurvProbAir}
P_{\rm Surv} = e^{-l /(c \tau) m ~ c^2 /E_\tau}
              %e^{-l (c \tau)^{-1} m c^2 E_\tau^{-1}}
\end{equation}

For a given $\tau$ flux exiting the Earth, we can estimate the detection 
rate, integrating in length the flux multiplied by both the differential 
decay probability and the probability of detecting the shower: 
\begin{equation}
\label{ShowerRate}
{dN_{sh} \over d E_\tau} = {dl \over c \tau} {dm c^2 \over E_\tau} 
e^{-l /(c \tau) m c^2 /E_\tau}
%e^{-l (c \tau)^{-1} m c^2 E_\tau^{-1}}
%^{-{l \over c \tau} {m c^2 \over E_\tau}} 
{\cal P}^\tau_{det}[l,E_{sh},\theta] \phi[E_\tau,\theta]
\end{equation}
We note that provided that ${\cal P}^\tau_{det}$ is known, 
the $l$ integral can be 
in principle performed independently of the $\tau$ flux and expressed 
as an overall probability factor for detection, to multiply the rate obtained 
in Eq.~\ref{Eq:TauFlux}. %in subsection~\ref{sec:eloss}. 
Unfortunately this requires detailed knowledge of the detector response and it 
is extremely dependent on the threshold behavior of the detector, 
which depends critically on shower development, geometry and the trigger 
conditions. The calculation of this is quite technical and difficult to 
estimate analytically. 
Precise calculations of ${\cal P}^\tau_{det}$ must be preformed for each
experiment. These functions are now being studied for detectors such as 
the surface array and fluorescence detector of the Pierre Auger 
Observatory, EUSO and others. 

To understand the effect of this factor for an air shower array  
we consider that the shower will be detected provided that the particle 
density exceeds a given threshold at the point when the plane of shower 
maximum intercepts ground level. The center of the shower in this plane 
lies at a distance $h = r_g \sin \theta$ to the ground level where $h$ 
is the altitude at which it lies above ground. 
The expression of $h$ is simple in terms of $l$ and $l_{sh}$:
\begin{equation}
\label{hofl}
h = \sqrt{R_{\oplus}^2 \cos^2 \theta + (l+l_{sh})^2 +
  2(l+l_{sh})R_{\oplus} \cos \theta}-R_{\oplus} 
\end{equation}
where $R_{\oplus}$ is the Earth's radius. 
For a given shower energy we calculate $r_{max}$, the lateral distance 
below which the electron density of the shower is above threshold $\rho_e$. 
We make $h=r_{max} \sin \theta$ and invert Eq.~\ref{hofl} to obtain the 
corresponding $l$. This value of $l$ which we will denote $l_{max}$ is the 
maximum distance the $\tau$ can travel before decaying to be successfully 
detected according to our simplifying assumptions. 

\begin{figure}[ht]
\centerline{
\mbox{\epsfig{figure=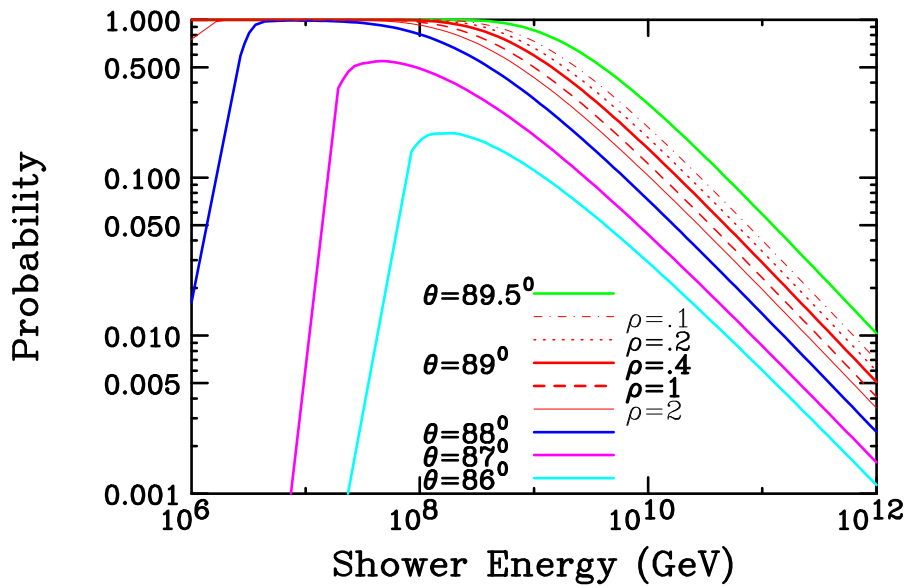,width=12.cm,angle=0}}
}
\caption{Integrated probability for the $\tau$ to decay in the atmosphere 
producing a shower that reaches ground level with electron number density 
($\rho_e$) above different threshold values.} 
\label{gndhitprob}
\end{figure}

%We will also consider that the shower is sufficiently 
%horizontal to ignore the altitude dependence of the atmospheric density. 
The function chosen for ${\cal P}^\tau_{det}$ allows analytic integration 
of the probability factor thus simplifying the calculation: 
\begin{equation}
\label{SurvProbAir}
{dN_{sh} \over d E_\tau} = 
{\cal P}_{det}[E_\tau,\theta] \phi[E_\tau,\theta]=
\left[ 1-e^{-l_{max} (c \tau)^{-1} m c^2 E_\tau^{-1}} \right] 
%e^{{l_{max} \over c\tau}{mc^2 \over E_\tau}} 
\phi[E_\tau,\theta]
\end{equation}
We have expressed the detection probability integrated over decay length 
as an overall factor to multiply the $\tau$ flux rate. 
The corresponding suppression factor has been plotted in 
Fig.~\ref{gndhitprob} for different zenith angles and different conditions 
to fix the minimum particle density to be detected at ground level. 
The probability is reduced for both low and high energy showers. Low energy 
showers do not meet the threshold requirement while high energy showers 
have maximum too high above ground. At energies around $10^8$~GeV 
the detection restriction becomes important even for showers 
with $87^{\circ}$ nadir angle. 

\subsection{The Resulting Shower Rate for an Air Shower Array}

The final shower rate recorded with an air shower array is severely 
constrained by the 
probability that the shower has sufficient particle density at ground level 
to be detected and identified as deep shower. This rate can be estimated 
taking the $\tau$ flux rate obtained in Eq.~\ref{Eq:TauFlux} and multiplying 
it by the integrated probability (as in Eq.~\ref{SurvProbAir}). 
We have calculated the resulting rates approximately as a function of the 
$\tau$ energy, taking into account that on average 
the $\tau$ decay showers carry about half the energy of the $\tau$. Different 
detection sensitivities are crudely implemented using $\rho_e=0.1,0.4$ and 
1~$m^{-2}$. 

\begin{figure}[ht]
\centerline{
\mbox{\epsfig{figure=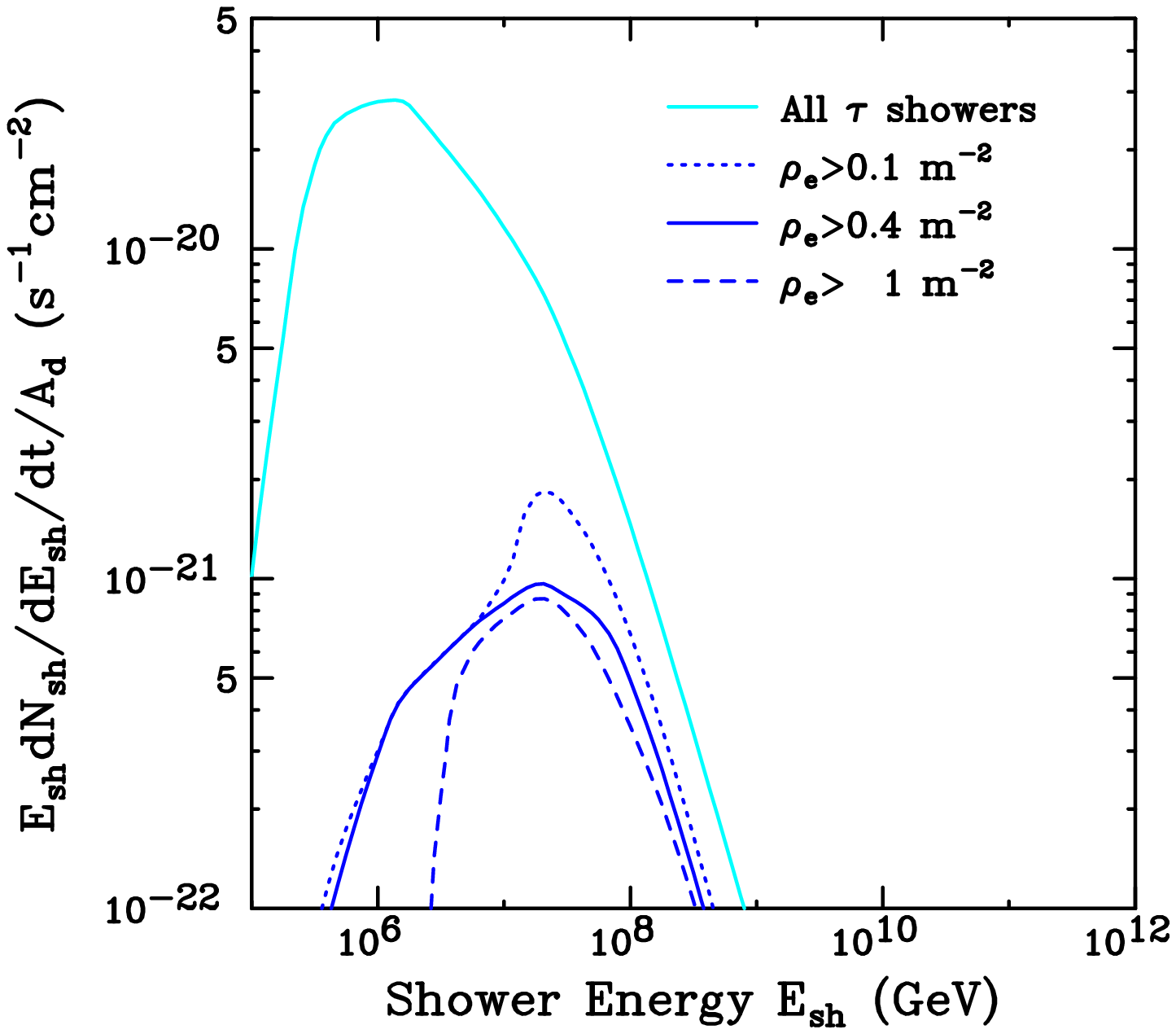,width=9.cm,angle=0},
\epsfig{figure=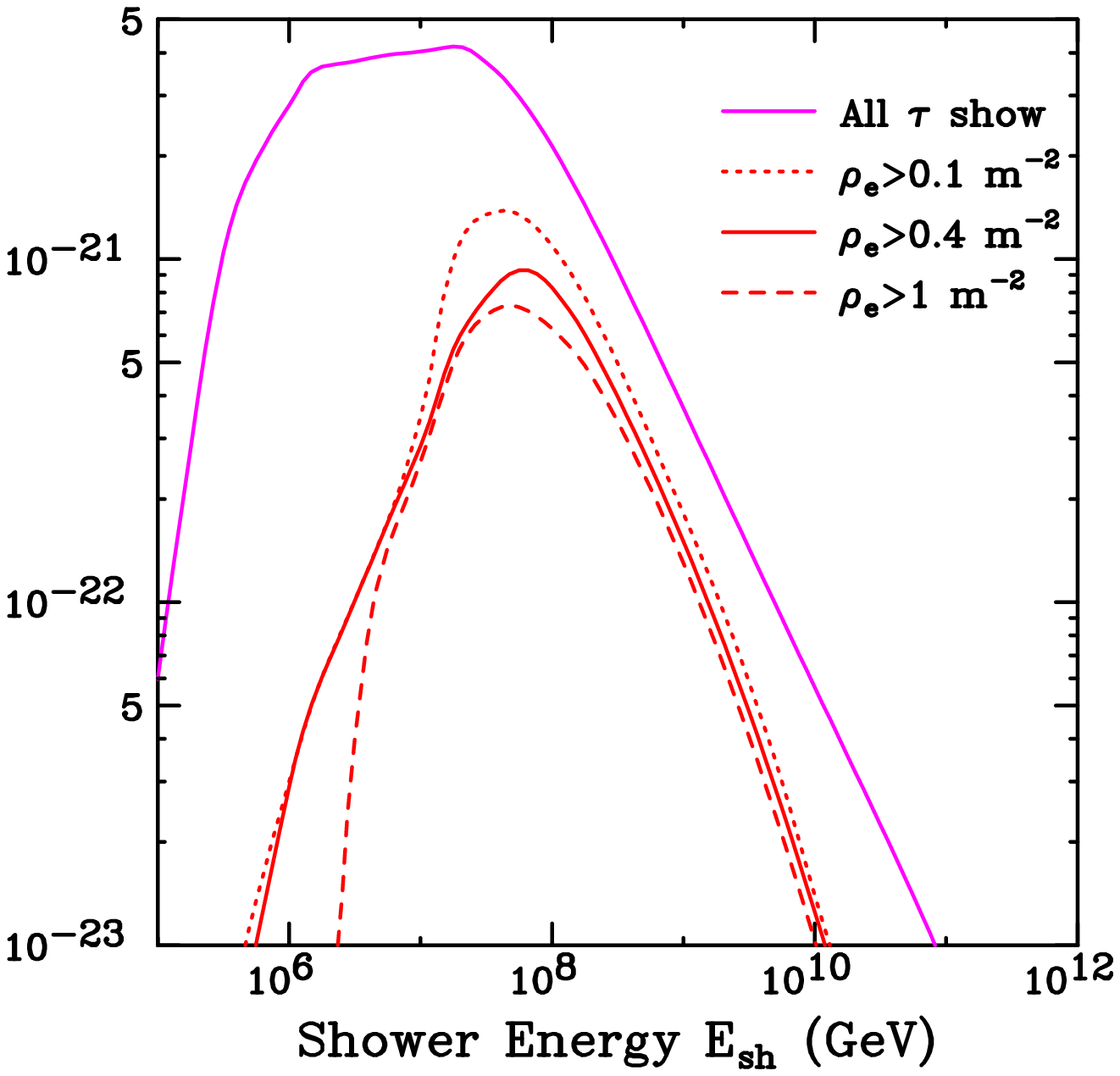,width=8.cm,angle=0}}
}
\caption{Comparison of the total $\tau$ shower rate produced by earth-skimming 
$\nu_\tau+\bar \nu_\tau$ flux assuming that all $\tau's$ make showers and the 
detected shower rate in 
an air shower array accounting for $\tau$ decay in the atmosphere and
requiring electron number density at ground to be above 
different thresholds ($\rho_e>$ 0.1,0.4,1~cm$^{-2}$ from top to bottom). 
The left panel is for the GRB flux and the right panel for the TD model.}
\label{ActualRate}
\end{figure}

The total shower rates for the two models chosen are compared in 
Figs.~\ref{ActualRate} to the actual rate when the shower is required to be 
detected by an air shower array. The total shower rates are just those shown in
the right hand panel of Fig.~\ref{TauFlux} rescaled to be differential in 
shower energy. 
When the decay probability and curvature of the Earth are accounted for as 
described in the previous subsection, the shower rate decreases quite
dramatically at low energies when compared to the production rate, 
as illustrated by the figure.  
At high energies the detection is suppressed simply because the showers 
decay too far away from the surface and this is not too dependent on $\rho_e$. 
At low energies detection details become crucial and constrain the detection 
in an intricate way. 
%

%\section{Comparisons with Other Calculations}
\section{Discussion}

\begin{figure}[ht]
\centerline{
\mbox{\epsfig{figure=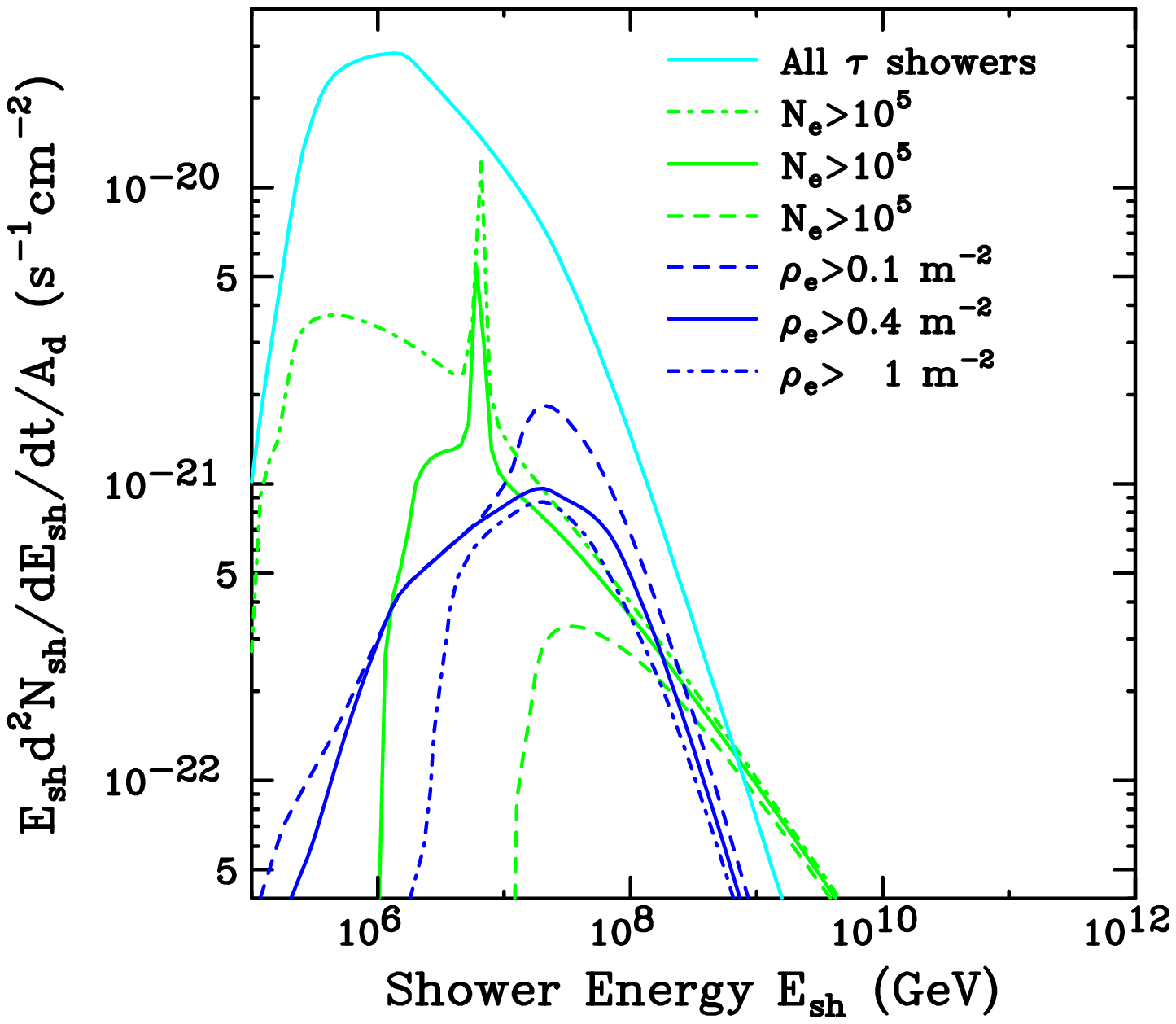,width=9.cm,angle=0},
\epsfig{figure=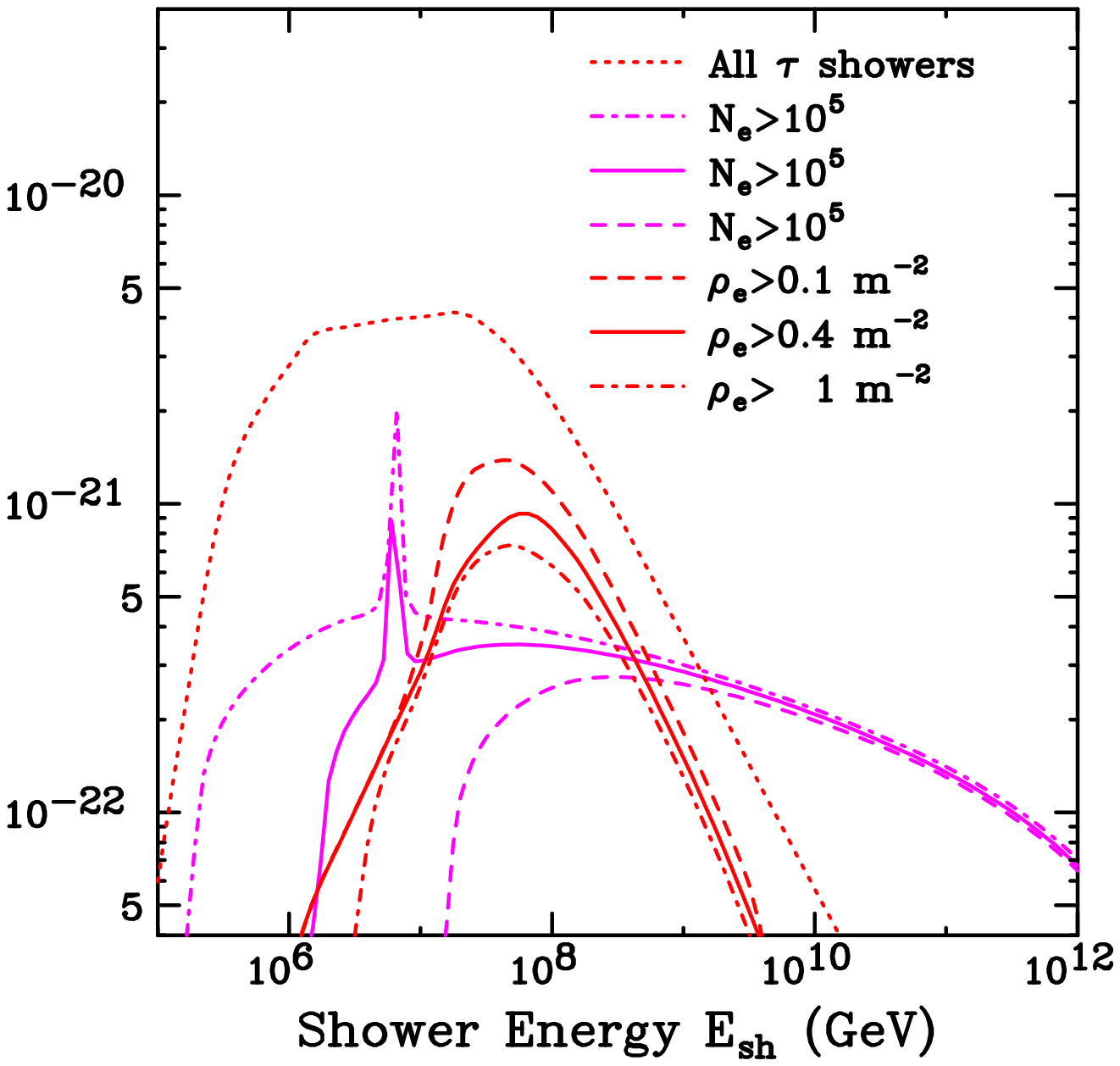,width=8.cm,angle=0}}
}
\caption{Down-going inclined shower rates compared to earth-skimming up-coming 
shower rates from $\tau$ decays. The plot is the same as Fig.~\ref{ActualRate} 
but it also includes the shower rates obtained for down-going 
neutrinos as shown in Fig.~\ref{ShRateTh}.} 
\label{CompRate}
\end{figure}

We have finally plotted the rates of down-going showers induced by neutrinos
compared to the up-coming shower rate induced by Earth-skimming neutrinos in 
Fig.~\ref{CompRate} for the two fluxes discussed in this article. The figure
compares the results of Figs.\ref{ShRateTh} and \ref{ActualRate}. 
In this figure the maximum earth-skimming $\tau$ shower rate (assuming all 
$\tau$'s induce a shower) is clearly shown to be larger than the down-going 
neutrino induced shower rate in the energy region between $10^5$-$10^9$~GeV. 
A fully efficient 1000~km$^2$ active area operating for a year corresponds to 
3~10$^{20}$s cm$^2$ of aperture. This is the 
order of magnitude required for detection of the $\tau$ produced 
by earth-skimming $\tau$ neutrinos for the lowest of the two discussed 
models (TD) assuming all $\tau$'s decay and are detected. 

Unfortunately it is not possible to detect all these $\tau$ decays with 
most existing detectors. Fluorescence detectors would be 
limited by duty cycle while detection of these showers by arrays of particle 
detectors on Earth surface is made complicated by an awkward geometry as 
indicated by 
the curves corresponding to the different thresholds. 
Moreover the energy region where the earth-skimming rates dominate is somewhat 
below the typical energies for detectors 
in construction such as the Auger detector. 
For a 1000~km$^2$ area the expected peak rate is of order 4 (14) events per year 
for the TD (GRB) model between $10^6$ and $4~10^7$~GeV ($3~10^5$ and 
$3~10^6$~GeV). This implies that if neutrinos are to be detected they would 
induce showers of energy very close to the detector threshold. 
As a result the rate calculations are quite uncertain unless very detailed
simulations are performed. Some of these have been made and results are 
in agreement with our calculations~\cite{bertou,miele}. 

The restriction introduced by air shower arrays somewhat compensates the 
duty factor for fluorescence detection and this explains why roughly 
similar rates can be obtained for the fluorescence and surface detector 
calculations of the earth skimming event rates. 

When decay probability, earth curvature and detector sensitivity are 
included, the obtained rates for down-going and earth skimming events are not 
very different for an air shower array. 
The differential shower rate from $\tau$ decay (induced by earth-skimming
neutrinos) dominates mostly in the energy 
region around $10^8$~GeV and is limited to an interval of about one and a 
half decades around this value. The relative rates will be very 
dependent on the detector sensitivity in this energy range. 
The potential for neutrino detection relies on the detector capability to 
identify very inclined showers which are almost horizontal in the case of 
$\nu_\tau$ earth skimming events. Once the zenith angle has been established 
the detector must also identify them as late development or deep showers. 
Although $\tau$ decay showers induced by earth-skimming neutrinos have  
higher potential for neutrino detection than down-going neutrino events, 
the actual relative rates in the air shower arrays is very dependent on
detector details and should be quite comparable for detectors having energy
thresholds in the $10^8$~GeV range.  

\ack 
I thank the Kavli Institute for Cosmological Physics at the University of 
Chicago for its hospitality, where part of this work was developed and most of
it written up. Much of the work and many ideas discussed here have been 
developed together with J. Alvarez Mu\~niz, M. Ave, J.J. Blanco Pillado, 
J.W. Cronin, 
F. Halzen, J. Hinton, A. Husain, G. Parente, T. Stanev, R.A. V\'azquez and 
A.A. Watson, over many years. I also thank P. Privitera for several 
discussions, A.A.~Watson for careful reading of the manuscript and 
Centro de Supercomputaci\'on de Galicia (CESGA) for computer resources. 
This work was supported in part by Ministerio de Educaci\'on y Ciencia 
(FPA 2001-3237, FPA 2002-01161, FPA 2004-01198), by Secretar\'\i a de Estado 
de Educaci\'on y Univesidades (Plan de Movilidad 2004), by Xunta de Galicia 
(PGIDIT02 PXIC 20611PN), by ESF neutrino network N-86, by Feder Funds and by 
the KICP under NSF grant PHY-0114422.


\section*{References}

\begin{thebibliography}{99}
%
\bibitem{Bird} D.J. Bird {\it et al.}, {\it Phys. Rev. Lett.} {\bf 71} (1993)
3401; D.J.~Bird {\it et al. Phys. Rev.} {\bf D}.
%
%\bibitem{halvaz} 
%F.~Halzen, R.A.~V\'azquez, T.~Stanev and CH.~Vankov, {\it Astropart. Phys.} 
%{\bf 3} (1995) 151-156. 
%
%\cite{Hayashida:1994hb}
\bibitem{Hayashida:1994hb}  N.~Hayashida {\it et al.},
  %``Observation of a very energetic cosmic ray well beyond the predicted 2.7-K
  %cutoff in the primary energy spectrum,''
  Phys.\ Rev.\ Lett.\  {\bf 73} (1994) 3491.
  %%CITATION = PRLTA,73,3491;%%
%
\bibitem{AGASA20} %\cite{Takeda:2002at} \bibitem{Takeda:2002at}
  M.~Takeda {\it et al.},
  %``Energy determination in the Akeno Giant Air Shower Array experiment,''
  Astropart.\ Phys.\  {\bf 19} (2003) 447
  [arXiv:astro-ph/0209422].
  %%CITATION = ASTRO-PH 0209422;%%
%
\bibitem{HiRes20} %\cite{Abbasi:2002ta} \bibitem{Abbasi:2002ta}
  R.~U.~Abbasi {\it et al.}  [High Resolution Fly's Eye Collaboration],
  %``Measurement of the flux of ultrahigh energy cosmic rays from monocular
  %observations by the High Resolution Fly's Eye experiment,''
  Phys.\ Rev.\ Lett.\  {\bf 92} (2004) 151101
  [arXiv:astro-ph/0208243].
  %%CITATION = ASTRO-PH 0208243;%%
%
\bibitem{GZK} Greisen K., {\it Phys. Rev. Lett.} {\bf 16} (1966) 748. 
%
%\cite{Zatsepin:1966jv}
\bibitem{Zatsepin:1966jv}
  G.~T.~Zatsepin and V.~A.~Kuzmin,
  %``Upper Limit Of The Spectrum Of Cosmic Rays,''
  JETP Lett.\  {\bf 4} (1966) 78
  [Pisma Zh.\ Eksp.\ Teor.\ Fiz.\  {\bf 4} (1966) 114].
  %%CITATION = JTPLA,4,78;%%
%
\bibitem{naganowatson} See for instance, 
M.~Nagano and A.~A.~Watson, Rev.\ Mod.\ Phys.\  {\bf 72}, 689 (2000), and
references therein. 
%``Observations and implications of the ultrahigh-energy cosmic rays,''
%
%\bibitem{zasmoriond} 
%E.~Zas, {\sl Proc. of  XXXIInd Rencontres de Moriond}, 
%"High-Energy Phenomena in Astrophysics", Les Arcs, Jan 1997, eds. 
%Giraud-Heraud, Y. and Tran Thanh Van, J. (Editions Frontieres, Paris), 
%pp 37-42 .
%%
\bibitem{bottom-up} For a review see for instance, 
A.V. Olinto, {\it Phys. Rept.} {\bf 333} (2000) 329; 
P. Bhattacharjee, G. Sigl, {\sl Phys. Rept.} {\bf 327} (2000) 109-247. 
%
%\cite{Gaisser:1994yf}
\bibitem{Nureview} For a review see for instance, 
  T.~K.~Gaisser, F.~Halzen and T.~Stanev,
  %``Particle astrophysics with high-energy neutrinos,''
  Phys.\ Rept.\  {\bf 258} (1995) 173
  [Erratum-ibid.\  {\bf 271} (1996) 355];
%  [arXiv:hep-ph/9410384]; 
  %%CITATION = HEP-PH 9410384;%%
%
%\cite{Learned:2000sw}
%\bibitem{Learned:2000sw}
  J.~G.~Learned and K.~Mannheim,
  %``High-energy neutrino astrophysics,''
  Ann.\ Rev.\ Nucl.\ Part.\ Sci.\  {\bf 50} (2000) 679; 
  %%CITATION = ARNUA,50,679;%%
%
%\cite{Halzen:2002pg}
%\bibitem{Halzen:2002pg}
  F.~Halzen and D.~Hooper,
  %``High-energy neutrino astronomy: The cosmic ray connection,''
  Rept.\ Prog.\ Phys.\  {\bf 65} (2002) 1025; 
%  [arXiv:astro-ph/0204527].
  %%CITATION = ASTRO-PH 0204527;%%
%
%\bibitem{Waxman04} 
E. Waxman, New J.\ Phys.\ {\bf 6} (2004) 140.
%
\bibitem{GZKnu}  F.W.~Stecker, C.~Done, M.H.~Salomon and P.~Sommers,
Phys.\ Rev.\ Lett.\ {\bf 66} (1991) 2697. 
%
\bibitem{berez} 
V.S.~Berezinsky and G.T.~Zatsepin {\sl Phys.\ Lett.\ }{\bf B28} (1969) 423; 
V.S. Berezinsky and A. Yu. Smirnov, {\sl Astrophys. Space Science} {\bf 32} 
(1975) 461.
%
\bibitem{Tokyo} T.~Hara {\sl et al.} 
{\sl Proc. of the XI Int. Cosmic Ray Conf.}, Budapest (1969), 
{\sl Acta Physica Academiae Scietiarum Hungaricae} {\bf 29}, Suppl. 3, p. 
361-367, (1970); Nagano,~M. {\sl et al.}, {\sl J.\ Phys. Soc. Japan}
M.~Nagano {\it et al.}, {\sl J. Phys. Soc. Japan.} {\bf 30} 
(1971) 33.
%
\bibitem{HP60s} A.M. Hillas {\sl et al.}, Proc. of the 11th ICRC, Budapest 
(1969), {\sl Acta Physica Academiae Scietiarum Hungaricae} {\bf 29}, Suppl. 3, 
pp. 533-538, (1970). D. Andrews {\sl et al.}, {\sl ibid} pp. 337-342, (1970).
Lawrence, M.A., Reid, R.J.O., and Watson, A.A. (1991). 
%
\bibitem{Kiraly} P. Kiraly {\it et al.}, {\sl J. Phys. A:Gen. Phys.} {\bf 4} 
(1971) 367.
%
\bibitem{tokyobremss} S.~Mikamo {\em et al.}, {\sl Lett.\ al Nuovo Cimento\ 
{\bf 34} N 10, (1982) 273}. 
%
\bibitem{hsvum}  E.~Zas, F.~Halzen and R.~A.~V\'azquez {\sl Astropart.
Phys.} {\bf 1} (1993) 297.
%
\bibitem{ParenteShoup} G.~Parente, A.~Shoup and G.B.~Yodh, {\sl Astropart.
Phys.}{\bf 3} (1995) 17.
%
\bibitem{AGASA} M. Nagano {\it et al.}, {\sl J. Phys. G: Nucl. Part. Phys.}
{\bf 12} (1986) 69.
%
\bibitem{hawaii}
%GIANT HORIZONTAL AIR SHOWERS: IMPLICATIONS FOR AGN NEUTRINO FLUXES.
F. Halzen and E. Zas, {\sl Phys. Lett.} {\bf B289} (1992) 184.
%
\bibitem{blancoPRL} J.J. Blanco-Pillado, R.A. V\'azquez, and E. Zas,
{\sl Phys. Rev. Lett.} {\bf 78} (1997) 3614.
%
\bibitem{Ave03}
M.~Ave, J.~A.~Hinton, R.~A.~Vazquez, A.~A.~Watson and E.~Zas,
{\sl Phys.\ Rev.} {\bf D67}, 043005 (2003). 
%
\bibitem{Auger} Design Report of the Pierre Auger Collaboration, October 1995.
%
\bibitem{ParenteZas} 
G.~Parente and E.~Zas, in {\sl Proceedings of the 7th Int.Symposium on 
Neutrino Telescopes.} p.~345, ed.\ by M.~Baldo Ceolin, Venice (1996); 
%
\bibitem{Capelle} 
K.~S.~Capelle, J.~W.~Cronin, G.~Parente and E.~Zas,
{\sl Astropart.\ Phys.\ } {\bf 8}, 321 (1998). 
%``On the detection of ultra high energy neutrinos with the Auger
%Observatory,'
%
\bibitem{euso} 
%\cite{Agnetta:2003ej} \bibitem{Agnetta:2003ej}
  G.~Agnetta {\it et al.}  [EUSO Collaboration],
  %``The ULTRA experiment: A supporting activity for the EUSO project,''
%\href{http://www.slac.stanford.edu/spires/find/hep/www?irn=5876265}{SPIRES entry}
{\it 28th International Cosmic Ray Conferences (ICRC 2003), Tsukuba, Japan, 31 Jul - 7 Aug 2003}
%
\bibitem{owl} 
%\cite{Stecker:2003wm} \bibitem{Stecker:2003wm}
  F.~W.~Stecker,
  %``Cosmic physics: The high energy frontier,''
  J.\ Phys.\ G {\bf 29} (2003) R47
  [arXiv:astro-ph/0309027].
  %%CITATION = ASTRO-PH 0309027;%%
%\cite{Fargion:1999se} \bibitem{Fargion:1999se}
\bibitem{fargion} 
  D.~Fargion, A.~Aiello and R.~Conversano,
  %``Horizontal tau air showers from mountains in deep valley: Traces of  UHECR
  %neutrino/tau,''
  arXiv:astro-ph/9906450.
  %%CITATION = ASTRO-PH 9906450;%%
%%\cite{Yeh:2004rp}
\bibitem{Yeh:2004rp}
  P.~Yeh {\it et al.}  [NuTel Collaboration],
  %``PeV cosmic neutrinos from the mountains,''
  Mod.\ Phys.\ Lett.\ A {\bf 19} (2004) 1117.
  %%CITATION = MPLAE,A19,1117;%%
\cite{Cao:2004sd}
\bibitem{Cao:2004sd}
  Z.~Cao, M.~A.~Huang, P.~Sokolsky and Y.~Hu,
  %``Ultra high energy nu/tau detection using cosmic ray tau neutrino telescope
  %used in fluorescence / Cerenkov light detection,''
  arXiv:astro-ph/0411677.
  %%CITATION = ASTRO-PH 0411677;%%%
\bibitem{oscildiscov} 
%\cite{Fukuda:2001nk} \bibitem{Fukuda:2001nk}
  S.~Fukuda {\it et al.}  [Super-Kamiokande Collaboration],
  %``Constraints on neutrino oscillations using 1258 days of  Super-Kamiokande
  %solar neutrino data,''
  Phys.\ Rev.\ Lett.\  {\bf 86} (2001) 5656
  [arXiv:hep-ex/0103033].
  %%CITATION = HEP-EX 0103033;%%
%
%\bibitem{maxmix} Maximal teta23 mixing
%
\bibitem{equalnus} 
%\cite{Athar:2000yw}\bibitem{Athar:2000yw}
  H.~Athar, M.~Jezabek and O.~Yasuda,
  %``Effects of neutrino mixing on high-energy cosmic neutrino flux,''
  Phys.\ Rev.\ D {\bf 62} (2000) 103007. 
%  [arXiv:hep-ph/0005104].
  %%CITATION = HEP-PH 0005104;%%
%
\bibitem{learned} 
%\cite{Learned:1994wg} \bibitem{Learned:1994wg}
  J.~G.~Learned and S.~Pakvasa,
  %``Detecting tau-neutrino oscillations at PeV energies,''
  Astropart.\ Phys.\  {\bf 3} (1995) 267
  [arXiv:hep-ph/9405296].
  %%CITATION = HEP-PH 9405296;%%
%
\bibitem{AtharParente} 
%\cite{Athar:2000rx} \bibitem{Athar:2000rx}
  H.~Athar, G.~Parente and E.~Zas,
  %``Prospects for observations of high-energy cosmic tau-neutrinos,''
  Phys.\ Rev.\ D {\bf 62} (2000) 093010.
%  [arXiv:hep-ph/0006123].
  %%CITATION = HEP-PH 0006123;%%
%
\bibitem{Saltzberg} 
%\cite{Halzen:1998be} \bibitem{Halzen:1998be}
  F.~Halzen and D.~Saltzberg,
  %``Tau neutrino appearance with a 1000-Megaparsec baseline,''
  Phys.\ Rev.\ Lett.\  {\bf 81} (1998) 4305. 
%  [arXiv:hep-ph/9804354].
  %%CITATION = HEP-PH 9804354;%%
%
%FIRST on Earth skimming taus? 
%\cite{Letessier-Selvon:2000kk}
\bibitem{antoine}
  A.~Letessier-Selvon,
  %``Establishing the GZK cutoff with ultra high energy tau neutrinos,''
  AIP Conf.\ Proc.\  {\bf 566} (2000) 157
  [arXiv:astro-ph/0009444].
  %%CITATION = ASTRO-PH 0009444;%%
%
%\cite{Stanev:1999ki}
%\bibitem{Stanev:1999ki}
%  T.~Stanev,
%  %``Possible tau appearance experiment with atmospheric neutrinos,''
%  Phys.\ Rev.\ Lett.\  {\bf 83} (1999) 5427 
%  [arXiv:astro-ph/9907018].
%  %%CITATION = ASTRO-PH 0407638;%%%\cite{Fargion:2000iz}
\bibitem{Fargion:2000iz}
  D.~Fargion,
  %``Discovering ultra high energy neutrinos by horizontal and upward tau
  %air-showers: First evidences in terrestrial gamma flashes,''
  Astrophys.\ J.\  {\bf 570}, 909 (2002)
  [arXiv:astro-ph/0002453] and Proc. of Int. Conf on Neutrino Telescopes, 
Venice 2003, vol. 2 pp. 433-455 [arXiv:hep-ph/0306238].
  %%CITATION = ASTRO-PH 0002453;%%
%\cite{Bertou:2001vm}
\bibitem{bertou}
  X.~Bertou, P.~Billoir, O.~Deligny, C.~Lachaud and A.~Letessier-Selvon,
  %``Tau neutrinos in the Auger observatory: A new window to UHECR sources,''
  Astropart.\ Phys.\  {\bf 17} (2002) 183
  [arXiv:astro-ph/0104452]. 
%
%\cite{Feng:2001ue}
\bibitem{Feng:2001ue}
  J.~L.~Feng, P.~Fisher, F.~Wilczek and T.~M.~Yu,
  %``Observability of earth-skimming ultra-high energy neutrinos,''
  Phys.\ Rev.\ Lett.\  {\bf 88} (2002) 161102
  [arXiv:hep-ph/0105067].
  %%CITATION = HEP-PH 0105067;%%
  %%CITATION = ASTRO-PH 0104452;%%
%
\bibitem{bottai} %\cite{Bottai:2002nn}\bibitem{Bottai:2002nn}
  S.~Bottai and S.~Giurgola,
  %``UHE and EHE neutrino induced taus inside the earth,''
  Astropart.\ Phys.\  {\bf 18} (2003) 539 
 [arXiv:astro-ph/0205325].
  %%CITATION = ASTRO-PH 0205325;%%
%\cite{Gupta:2002ze}
\bibitem{Gupta:2002ze}
  N.~Gupta,
  %``The appearance of tau neutrinos from a gamma ray burst,''
  Phys.\ Lett.\ B {\bf 541} (2002) 16 [arXiv:astro-ph/0205451].
%\bibitem{Gupta:2003mz}
%  N.~Gupta,
  %``A study on the appearance of tau neutrinos from a gamma ray burst by
  %detecting their horizontal electromagnetic showers,''
and Phys.\ Rev.\ D {\bf 68} (2003) 063006.
%  [arXiv:astro-ph/0306007].
  %%CITATION = ASTRO-PH 0306007;%%
  %%CITATION = ASTRO-PH 0205451;%%
%\cite{Tseng:2003pn}
\bibitem{Tseng:2003pn}
  J.~J.~Tseng, T.~W.~Yeh, H.~Athar, M.~A.~Huang, F.~F.~Lee and G.~L.~Lin,
  %``The energy spectrum of tau leptons induced by the high energy
  %Earth-skimming neutrinos,''
  Phys.\ Rev.\ D {\bf 68} (2003) 063003
  [arXiv:astro-ph/0305507].
  %%CITATION = ASTRO-PH 0305507;%%
%\cite{Athar:2003nc}
\bibitem{Athar:2003nc}
  H.~Athar, K.~Cheung, G.~L.~Lin and J.~J.~Tseng,
  %``The high-energy galactic tau neutrino flux and its atmospheric
  %background,''
  Eur.\ Phys.\ J.\ C {\bf 33} (2004) S959
  [arXiv:astro-ph/0311586].
  %%CITATION = ASTRO-PH 0311586;%%
%
%\cite{Aramo:2004pr}
\bibitem{miele}
  C.~Aramo, A.~Insolia, A.~Leonardi, G.~Miele, L.~Perrone, O.~Pisanti and D.~V.~Semikoz,
  %``Earth-skimming UHE tau neutrinos at the fluorescence detector of Pierre
  %Auger Observatory,''
  Astropart.\ Phys.\  {\bf 23} (2005) 65
  [arXiv:astro-ph/0407638].
%
\bibitem{Han04} 
%\cite{Han:2004kq} \bibitem{Han:2004kq}
  T.~Han and D.~Hooper,
  %``The particle physics reach of high-energy neutrino astronomy,''
  New J.\ Phys.\  {\bf 6} (2004) 150
  [arXiv:hep-ph/0408348].
  %%CITATION = HEP-PH 0408348;%%
%
\bibitem{hires}
R.~U.~Abbasi {\it et al.} [High Resolution Fly's Eye Collaboration],
Phys.\ Rev.\ Lett.\  {\bf 92}, 151101 (2004);
Astrop.\ Phys.\  {\bf 22}, 139 (2004). 
%``Measurement of the flux of ultrahigh energy cosmic rays from monocular
%observations by the High Resolution Fly's Eye experiment,''
%\bibitem{Abbasi:2004dx}
%R.~U.~Abbasi {\it et al.}  [The High Resolution Fly's Eye Collaboration
%                  (HiRes)],
%``A Search for Arrival Direction Clustering in the HiRes-I monocular data 
%above
%10**19.5 eV,''
%
\bibitem{agasa}M.~Takeda {\it et al.} % [AGASA Collaboration],
Astropart.\ Phys.\  {\bf 19}, 135 (2003).
%``The arrival direction distribution of extremely high energy cosmic rays
%observed by AGASA,''
%\href{http://www.slac.stanford.edu/spires/find/hep/www?irn=5903084}{SPIRES 
%entry}
%{\it 28th International Cosmic Ray Conferences (ICRC 2003), 
%Tsukuba, Japan,} vol 1, pp. 437; 381.
%
%
\bibitem{EngArray} 
J.~Abraham {\it et al.} % [Pierre Auger Collaboration],
%``Properties and performance of the prototype instrument for the Pierre Auger
%Observatory,''
{\sl Nucl.\ Instrum.\ Meth.}\ {\bf A 523}, 50 (2004).
%
\bibitem{HSmodel} M.~Ave, R.A.~V\'azquez, and E.~Zas,
{\sl Astropart.\ Phys.\ } {\bf 14} (2000) 91.
%
\bibitem{rateave98}
M.~Ave, R.~A.~Vazquez, E.~Zas, J.~A.~Hinton and A.~A.~Watson, %{\sl et al.}, 
{\sl Astropart.\ Phys.\ }  {\bf 14}, 109 (2000).
%``The rate of cosmic ray showers at large zenith angles: A step towards  the
%detection of ultra-high energy neutrinos by the Pierre Auger  Observatory,''
%
%
\bibitem{AvePRL} M. Ave, J.A. Hinton, R.A. Vazquez, A.A. Watson,
and E. Zas, Phys. Rev. Lett. {\bf 85}, (2000) 2244.
%%\cite{Ave:2000nd} \bibitem{Ave:2000nd}
%  M.~Ave, J.~A.~Hinton, R.~A.~Vazquez, A.~A.~Watson and E.~Zas,
  %``New constraints from Haverah Park data on the photon and iron fluxes of
  %UHE cosmic rays,''
%  Phys.\ Rev.\ Lett.\  {\bf 85} (2000) 2244
%  [arXiv:astro-ph/0007386].
  %%CITATION = ASTRO-PH 0007386;%%
%
\bibitem{Ave00}
M.~Ave, J.~A.~Hinton, R.~A.~Vazquez, A.~A.~Watson and E.~Zas,
Phys.\ Rev.\ Lett.\  {\bf 85}, 2244 (2000).
%
\bibitem{billoir} 
%\cite{Billoir:1999nq}\bibitem{Billoir:1999nq}
  P.~Billoir,
  %``Neutrino capabilities of the Auger detector,''
%\href{http://www.slac.stanford.edu/spires/find/hep/www?irn=4288475}{SPIRES entry}
{\it 8th International Workshop on Neutrino Telescopes, Venice, Italy, 23-26
  Feb 1999},  vol. 2 pp. 111-122.
%
\bibitem{LCB} L.~Cazon, R.~A.~Vazquez, A.~A.~Watson and E.~Zas, 
{\sl Astropart.\ Phys.\ }  {\bf 21}, 71 (2004). 
%``Time structure of muonic showers,'' %{\sl et al.}, 
%
\bibitem{CroninFADC} J.W.~Cronin, {\sl Nucl.\ Phys.\ } (Proc. Suppl.) 
{\bf B 138}, 465 (2005).
%
\bibitem{avetokyo} M. Ave (Auger Colab.) {\it 28th 
International Cosmic Ray Conference (ICRC 2003), Tsukuba, Japan,}
vol 1, pp. 365.
%
%\cite{Mccomb:1982nz}
\bibitem{Mccomb:1982nz}
  T.~J.~L.~Mccomb and K.~E.~Turver,
  %``Imagine Of The Muon Cascade Of Large Extensive Air Showers,''
  J.\ Phys.\ Soc.\ Jap.\  {\bf 51} (1982) 3087.
%
\bibitem{LCB2} L.~Cazon, R.~A.~Vazquez and E.~Zas, 
(in press.) {\sl Astropart.\ Phys.\ }  (2005); arXiv:astro-ph/0412338. 
%\cite{Cazon:2004zx}
%\bibitem{Cazon:2004zx}
%  L.~Cazon, R.~A.~Vazquez and E.~Zas,
  %``Depth development of extensive air showers from muon time distributions,''
%  arXiv:astro-ph/0412338.
  %%CITATION = ASTRO-PH 0412338;%%
%
%\bibitem{EeVMuStanev} 
%
\bibitem{Gondolo} 
%\cite{Gelmini:2002sw} \bibitem{Gelmini:2002sw}
  G.~Gelmini, P.~Gondolo and G.~Varieschi,
  %``Measuring the prompt atmospheric neutrino flux with down-going muons in
  %neutrino telescopes,''
  Phys.\ Rev.\ D {\bf 67} (2003) 017301.
%  [arXiv:hep-ph/0209111].
  %%CITATION = HEP-PH 0209111;%%
%
\bibitem{Gonzalez} M.C.~Gonz\'alez-Garc\'\i a, F.~Halzen, R.A.~V\'azquez and
E.~Zas, {\sl Phys. Rev.} {\bf D 49} (1994) 2310.
%
\bibitem{PZaugerMuons} G.~Parente and E.~Zas, {\sl Proc. XXV Int.
 Cosmic Ray Conf.} (Durban 1997), vol {\bf 7} pp. 201. 
%
%\bibitem{Ffits} Structure function fits 
%%\cite{Dutta:2002zc}
%
\bibitem{Reno} R. Gandhi {\it et al.}, {\sl Astropart. Phys.} {\bf 5 } (1996) 
81. Glashow resonance? 
%
\bibitem{Parente} Contribution EPS 0217 to the Int. Europhys. Conf. on High
 Energy Physics (HEP 95), Brussels; {\sl Proc. XXV Int.
 Cosmic Ray Conf.} (Durban 1997), vol {\bf 7}, pp.~109-112.
%
%\cite{Gandhi:1998ri}
\bibitem{Gandhi:1998ri}
  R.~Gandhi, C.~Quigg, M.~H.~Reno and I.~Sarcevic,
  %``Neutrino interactions at ultrahigh energies,''
  Phys.\ Rev.\ D {\bf 58} (1998) 093009
  [arXiv:hep-ph/9807264].
  %%CITATION = HEP-PH 9807264;%%
%\cite{Kretzer:2002fr}
%
\bibitem{Kretzer:2002fr}
  S.~Kretzer and M.~H.~Reno,
  %``Tau neutrino deep inelastic charged current interactions,''
  Phys.\ Rev.\ D {\bf 66} (2002) 113007
  [arXiv:hep-ph/0208187].
  %%CITATION = HEP-PH 0208187;%%
% mass corrections of tau reduce xs few percent!
%
\bibitem{Reno:2004cx}
  M.~H.~Reno,
  %``High energy neutrino cross sections,''
  arXiv:hep-ph/0410109.
  %%CITATION = HEP-PH 0410109;%%
%
\bibitem{castro2} 
%\cite{CastroPena:2000sx} \bibitem{CastroPena:2000sx}
  J.~A.~Castro Pena, G.~Parente and E.~Zas,
  %``Nuclear effects on the UHE neutrino nucleon deep inelastic scattering
  %cross section,''
  Phys.\ Lett.\ B {\bf 507} (2001) 231.
%  [arXiv:hep-ph/0011309].
%%CITATION = HEP-PH 0011309;%%
%
\bibitem{CTEQ6} 
%\cite{Pumplin:2002vw}\bibitem{Pumplin:2002vw}
  J.~Pumplin, D.~R.~Stump, J.~Huston, H.~L.~Lai, P.~Nadolsky and W.~K.~Tung,
  %``New generation of parton distributions with uncertainties from global  QCD
  %analysis,''
  JHEP {\bf 0207} (2002) 012
  [arXiv:hep-ph/0201195].
  %%CITATION = HEP-PH 0201195;%%
%
%\cite{CastroPena:2000fx}\bibitem{CastroPena:2000fx}
\bibitem{castro} J.~A.~Castro Pena, G.~Parente and E.~Zas,
  %``Measuring the BFKL pomeron in neutrino telescopes,''
  Phys.\ Lett.\ B {\bf 500} (2001) 125
  [arXiv:hep-ph/0011043].
  %%CITATION = HEP-PH 0011043;%%
%
\bibitem{alz00} J.~Alvarez-Mu\~niz, R.A.~V\'azquez, E.~Zas, 
{\sl Phys. \ Rev.}  {\bf D 61}, 023001 (2000).
%
\bibitem{WeilerXS} %T. Weiler X-section
%\cite{Weiler:2003ud} \bibitem{Weiler:2003ud}
  T.~J.~Weiler,
  %``Physics with cosmic neutrinos, PeV to ZeV,''
  Int.\ J.\ Mod.\ Phys.\ A {\bf 18} (2003) 4065.
%  [arXiv:astro-ph/0304180].
%%CITATION = ASTRO-PH 0304180;%%
\bibitem{GRBflux} %\cite{Waxman:1997ti} \bibitem{Waxman:1997ti}
  E.~Waxman and J.~N.~Bahcall,
  %``High energy neutrinos from cosmological gamma-ray burst fireballs,''
  Phys.\ Rev.\ Lett.\  {\bf 78} (1997) 2292; 
%  [arXiv:astro-ph/9701231]
  %%CITATION = ASTRO-PH 9701231;%%
%\cite{Razzaque:2004yv}
%\bibitem{Razzaque:2004yv}
  S.~Razzaque, P.~Meszaros and E.~Waxman,
  %``TeV neutrinos from core collapse supernovae and hypernovae,''
  Phys.\ Rev.\ Lett.\  {\bf 93} (2004) 181101.
 % [arXiv:astro-ph/0407064].
  %%CITATION = ASTRO-PH 0407064;%%
%
\bibitem{TDFlux}  %\cite{Semikoz:2003wv}
%\cite{Protheroe:1996pd} \bibitem{Protheroe:1996pd}
  R.~J.~Protheroe and T.~Stanev,
  %``Limits on models of the ultrahigh energy cosmic rays based on  topological
  %defects,''
  Phys.\ Rev.\ Lett.\  {\bf 77} (1996) 3708
  [Erratum-ibid.\  {\bf 78} (1997) 3420]
  [arXiv:astro-ph/9605036].
  %%CITATION = ASTRO-PH 9605036;%%
\bibitem{Semikoz:2003wv}
  D.~V.~Semikoz and G.~Sigl,
  %``Ultra-high energy neutrino fluxes: New constraints and implications,''
  JCAP {\bf 0404} (2004) 003
  [arXiv:hep-ph/0309328].
  %%CITATION = HEP-PH 0309328;%%
%
\bibitem{Gaisser} T.K.~Gaisser, Cosmic Rays and Particle Physics, Cambridge
Univ. Press, 1990. %; and references therein.
%
\bibitem{greisen} K.~Greisen, Prog. in Cosmic Ray Phys., ed. J.G. Wilson,
{\bf Vol.III}, p.1, North Holland Publ. Co. (1956).
%
\bibitem{PZAuger05} G.~Parente and E.~Zas in preparation.
%
\bibitem{RegDutta} 
% \bibitem{Dutta:2002im}
  S.~I.~Dutta, M.~H.~Reno and I.~Sarcevic,
  %``High-energy tau neutrinos,''
  AIP Conf.\ Proc.\  {\bf 624} (2002) 271; 
  %%CITATION = APCPC,624,271;%%
%
%\cite{Jones:2004ir}
%\bibitem{Jones:2004ir}
  J.~Jones, I.~Mocioiu, I.~Sarcevic and M.~H.~Reno,
  %``Tracing very high energy tau neutrinos from cosmological sources in ice,''
  arXiv:hep-ph/0408060.
  %%CITATION = HEP-PH 0408060;%%
%\cite{Dutta:2000hh}
%
\bibitem{Dutta:2000hh}
  S.~I.~Dutta, M.~H.~Reno, I.~Sarcevic and D.~Seckel,
  %``Propagation of muons and taus at high energies,''
  Phys.\ Rev.\ D {\bf 63} (2001) 094020
  [arXiv:hep-ph/0012350].
  %%CITATION = HEP-PH 0012350;%%
%
\bibitem{Lipari} P.~Lipari and T.~Stanev, {\sl Phys.\ Rev.\ } 
{\bf D44} (1991) 3543.
%
\bibitem{haverah} R.M. Tennent, {\sl Proc Phys Soc} {\bf 92} 
(1967) 622. M.A. Lawrence, R.J.O. Reid, and A.A. 
Watson, {\sl J Phys G} {\bf 17} (1991) 733. 
%
\bibitem{physrep} T.K. Gaisser, F. Halzen and T. Stanev, Phys. Rep.
238 (1995) 173.
%

\end{thebibliography}
\end{document}